\documentclass[twocolumn, twocolappendix, times]{aastex63}
\usepackage{amsmath}
\usepackage{graphicx}

\received{*}
\revised{*}
\accepted{*}
\submitjournal{ApJ}
\shorttitle{Stellar Superflares Observed Simultaneously with Kepler and XMM-Newton}
\shortauthors{Kuznetsov and Kolotkov}

\begin{document}
\title{Stellar Superflares Observed Simultaneously with \textit{Kepler} and \textit{XMM-Newton}}

\correspondingauthor{Alexey A. Kuznetsov}
\email{a\_kuzn@iszf.irk.ru}

\author[0000-0001-8644-8372]{Alexey A. Kuznetsov}
\affiliation{Institute of Solar-Terrestrial Physics, Irkutsk, 664033, Russia}

\author[0000-0002-0687-6172]{Dmitrii Y. Kolotkov}
\affiliation{Centre for Fusion, Space and Astrophysics, Department of Physics, University of Warwick, CV4 7AL, UK}
\affiliation{Institute of Solar-Terrestrial Physics, Irkutsk, 664033, Russia}

\begin{abstract}
Solar and stellar flares are powerful events which produce intense radiation across the electromagnetic spectrum. Multiwavelength observations are highly important for understanding the nature of flares, because different flare-related processes reveal themselves in different spectral ranges. To study the correlation between thermal and nonthermal processes in stellar flares, we have searched the databases of \textit{Kepler} (optical observations) and \textit{XMM-Newton} (soft X-rays) for the flares observed simultaneously with both instruments; nine distinctive flares (with energies exceeding $10^{33}$ erg) on three stars (of K-M spectral classes) have been found. We have analyzed and compared the flare parameters in the optical and X-ray spectral ranges; we have also compared the obtained results with similar observations of solar flares. Most of the studied stellar flares released more energy in the optical range than in X-rays. In one flare, X-ray emission strongly dominated, which could be caused either by soft spectrum of energetic electrons or by a near-limb position of this flare. The X-ray flares were typically delayed with respect to and shorter than their optical counterparts, which is partially consistent with the Neupert effect. Using the scaling laws based on the magnetic reconnection theory, we have estimated the characteristic magnetic field strengths in the stellar active regions and the sizes of these active regions as about $25-70$ G and $250\,000-500\,000$ km, respectively. The observed stellar superflares appear to be scaled-up versions of solar flares, with a similar underlying mechanism and nearly the same characteristic magnetic field values, but with much larger active region sizes.
\end{abstract}


\section{Introduction}
Flares on the Sun and other stars are caused by a fast explosive release of magnetic energy stored in the solar or stellar corona \citep{Haisch1991}. According to the ``standard'' scenario \citep[e.g.,][]{Benz2010}, the magnetic reconnection and energy release occur high in the corona and result in a highly effective acceleration of electrons up to relativistic energies. These nonthermal electrons then propagate downward causing the chromosphere heating and chromospheric evaporation; both the nonthermal particles and hot thermal plasma produce intense electromagnetic radiation at different levels of the solar/stellar atmosphere and in different wavelength ranges---from radio to $\gamma$-rays.

Many stars have been found to produce very powerful flares (``superflares''), with the released energies exceeding the energies of the largest known solar flares by several orders of magnitude. The superflares have been detected, in particular, on the RS CVn binaries, young T Tauri stars, and UV Ceti-like red dwarfs \citep{Haisch1991}, as well as on the more or less solar-like G dwarfs \citep[][etc.]{Maehara2012}. Studying the stellar superflares offers opportunities for better understanding the origin and physical mechanisms of flares in general, and also for answering an intriguing question about the possibility (and probability) of similar extreme events on our Sun \citep[][etc.]{Shibata2013}; in addition, superflares are considered to be a significant factor affecting habitability of exoplanets \citep[e.g.,][]{Lingam2017}.

Studying the solar flares benefits greatly from the available simultaneous multiwavelength observations, which allow one to investigate the processes at different levels of the atmosphere and to reconstruct a more comprehensive picture of the phenomenon. At the same time, multiwavelength observations of stellar flares are more difficult to arrange and hence much less common. In this work, we analyze the simultaneous observations of stellar flares with \textit{Kepler} \citep[in the optical range;][]{Borucki2010} and \textit{XMM-Newton} \citep[in the soft X-ray range;][]{Jansen2001}. The white-light continuum emission of the solar and stellar flares is believed to be mostly a blackbody radiation from the regions of the chromosphere (and, probably, upper photosphere) heated by nonthermal ($\gtrsim 50$ keV) electron beams \citep[][etc.]{Neidig1989, Benz2010}. On the other hand, the soft X-rays represent mostly a thermal ($\sim 10$ MK) bremsstrahlung radiation from hot plasma in the coronal flaring loops \citep[][etc.]{Gudel2004, Benz2010}. Therefore simultaneous observations in the optical and soft X-ray ranges allow one to investigate correlations between the thermal and nonthermal processes in flares. The multiwavelength observations are also necessary to estimate the total flare energetics, which, in turn, constrains the characteristics of the magnetic reconnection process. 

In this work, we search the \textit{Kepler} and \textit{XMM-Newton} databases for the stellar flares observed simultaneously by both instruments. We analyze the \textit{Kepler} and \textit{XMM-Newton} light curves using the Markov chain Monte Carlo approach to determine the flare parameters; we estimate and compare the flare luminosities and energies in both spectral ranges. We also compare the obtained results with similar observations of solar flares.

\section{Data and sample selection}
\subsection{Matching the \textit{Kepler} and \textit{XMM-Newton} databases}
Simultaneous \textit{Kepler} and \textit{XMM-Newton} observations of superflares on the young Pleiades stars during the \textit{Kepler}/K2 mission have been studied earlier by \citet{Guarcello2019a, Guarcello2019b}; in contrast, we consider here the primary \textit{Kepler} observational campaign (May 2009 -- May 2013). Simultaneous \textit{Kepler} and \textit{XMM-Newton} observations during the primary \textit{Kepler} campaign have been studied before by \citet{Pizzocaro2019}; however, \citet{Pizzocaro2019} focused mainly on the general stellar activity indicators (average X-ray luminosity, flare occurrence rate, etc.) and did not analyze the individual flares in detail.

\begin{deluxetable}{lccc}
\renewcommand{\tabcolsep}{4pt}
\tablewidth{0pt}
\tablecaption{Parameters of the selected stars: spectral types (SpT), effective temperatures ($T_{\mathrm{eff}}$), luminosities ($L$), distances ($d$), metallicities ([Fe/H]), rotation periods ($P_{\mathrm{rot}}$), optical extinctions ($A_V$), and magnitudes in the Johnson ($B$ and $V$), 2MASS ($K_S$) and \textit{Gaia} ($G$, $G_{\mathrm{BP}}$ and $G_{\mathrm{RP}}$) bands. We present also the estimated component masses ($M_{\mathrm{star}}$) and radii ($R_{\mathrm{star}}$) and orbital separations ($a$) for the supposed binaries (see Section \ref{BinarySection}).\label{stars}}
\tablehead{Star & \colhead{KIC 8093473} & \colhead{KIC 8454353} & \colhead{KIC 9048551}}
\startdata
SpT\,$^a$ & M3 & M2 & K7\\
$T_{\mathrm{eff}}$, K\,$^b$ & $3357_{3344}^{3528}$ & $3541_{3514}^{3923}$ & $4124_{4025}^{4138}$\\
$L$, $L_{\sun}$\,$^b$ & $0.111_{0.103}^{0.120}$ & $0.051_{0.050}^{0.052}$ & $0.084_{0.084}^{0.085}$\\
$d$, pc\,$^b$ & $205.9_{198.4}^{213.9}$ & $168.5_{167.5}^{169.4}$ & $125.9_{125.6}^{126.2}$\\ 
\textrm{[Fe/H]}, dex & $+0.04_{-0.06}^{+0.14}\,^c$ & $-0.16_{-0.27}^{-0.05}\,^c$ & $-0.04_{-0.05}^{-0.03}\,^d$\\
$P_{\mathrm{rot}}$, days\,$^e$ & 6.043 & 1.496 & 8.553\\
$A_V$, mag\,$^f$ & 0.171 & 0.109 & 0.080\\
\hline
$B$, mag\,$^g$ & 17.273 & 17.297 & 15.345\\
$V$, mag\,$^g$ & 15.883 & 16.005 & 14.085\\
$K_S$, mag\,$^h$ & $11.160_{11.149}^{11.171}$ & $11.551_{11.531}^{11.571}$ & $10.466_{10.448}^{10.484}$\\
$G$, mag\,$^b$ & $14.696_{14.694}^{14.698}$ & $14.941_{14.940}^{14.943}$ & $13.336_{13.335}^{13.337}$\\
$G_{\mathrm{BP}}$, mag\,$^b$ & $15.989_{15.980}^{15.999}$ & $16.094_{16.086}^{16.101}$ & $14.187_{14.181}^{14.192}$\\
$G_{\mathrm{RP}}$, mag\,$^b$ & $13.561_{13.556}^{13.566}$ & $13.863_{13.859}^{13.867}$ & $12.439_{12.434}^{12.444}$\\
\hline
$M_{\mathrm{star}}$, $M_{\sun}$\,$^i$ & 0.57 & 0.44 & \nodata\\
$R_{\mathrm{star}}$, $R_{\sun}$\,$^i$ & 0.55 & 0.42 & \nodata\\
$a$, AU\,$^i$ & 0.068 & 0.024 & \nodata\\
\enddata
\tablerefs{$^{(a)}$Spectral types were estimated from the effective temperatures according to \citet{Pecaut2013}; $^{(b)}$\citet{Gaia2018}; $^{(c)}$\citet{Gaidos2016}; $^{(d)}$\citet{Jonsson2020}; $^{(e)}$\citet{McQuillan2014}; $^{(f)}$\citet{Brown2011}; $^{(g)}$\citet{Pizzocaro2019}; $^{(h)}$\citet{Cutri2003}; $^{(i)}$this work.}
\end{deluxetable}

The \textit{Kepler} data archive\footnote{\url{https://archive.stsci.edu/kepler}} provides nearly con\-ti\-nu\-o\-us light curves for hundreds of thousands of stars within its field of view, for the above-mentioned time range. As the starting point for the X-ray data, we used the 3XMM-DR5 serendipitous source catalog \citep{Rosen2016}, which contains hundreds of thousands of sources detected by \textit{XMM-Newton}; light curves are provided for the brightest sources (including the signals from the selected source regions and from nearby background regions, to filter out the background fluctuations). To find the flares that occurred simultaneously in both spectral ranges, we (a) selected the \textit{XMM-Newton} (3XMM-DR5) detections within the \textit{Kepler} primary campaign field of view and time range; (b) matched the \textit{Kepler} Input Catalog \citep[KIC,][]{Brown2011} and the 3XMM-DR5 catalog to select the objects with the mutual positional differences of $\lesssim 5''$ \citep[this value is determined by the 3XMM-DR5 positional error, cf.][while the KIC positional error is negligible]{Rosen2016}, with account for the proper motion; (c) selected the \textit{XMM-Newton} (3XMM-DR5) detections with $>10^3$ source counts in any EPIC detector ($0.2-12$ keV range), for which reliable X-ray light curves with a sufficiently high time resolution are available \citep[see][]{Rosen2016}; (d) inspected the \textit{Kepler} and \textit{XMM-Newton} light curves of the selected objects visually, to search for simultaneous flares. As a result, we have identified three stars that exhibited well-defined correlated peaks in the optical and X-ray light curves\footnote{Our list of stars with simultaneous optical and X-ray flares overlaps partially with that in \citet{Pizzocaro2019}, but is not exactly the same due to different selection criteria.}; parameters of these stars are summarized in Table \ref{stars}.

\begin{figure}
\centerline{\includegraphics{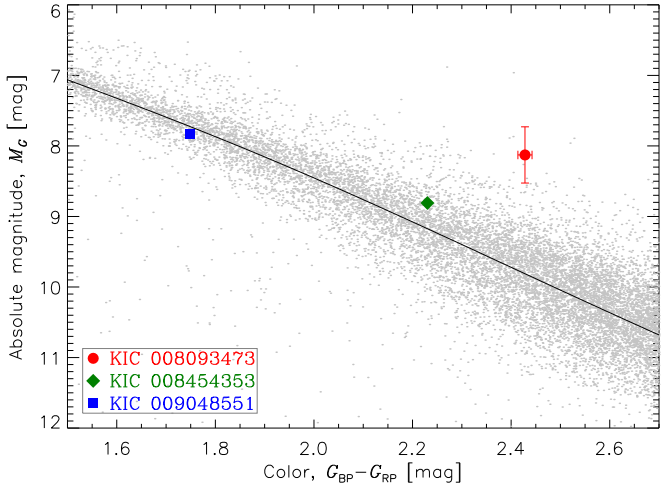}}
\caption{Locations of the selected stars (see Table \protect\ref{stars}) on the Hertzsprung-Russell diagram, with the absolute stellar magnitude in the \textit{Gaia} band $M_G$ plotted vs. the \textit{Gaia} $G_{\mathrm{BP}}-G_{\mathrm{RP}}$ color. Gray dots represent \textit{Gaia} stars within 50 pc distance; the solid line is the corresponding median main sequence. The error bars correspond to $1\sigma$ level; for KIC 8454353 and KIC 9048551, the error bars are nearly the same as the symbol sizes.}
\label{HRdiagram}
\end{figure}

\subsection{Parameters of the selected stars}
To check further the stellar parameters for our sample, we have analyzed the available catalogs and photometry data; we have compared the observations with theoretical PARSEC isochrones \citep{Marigo2017} and gyrochronology relations by \citet{Barnes2007}, \citet{Mamajek2008}, and \citet{Angus2019}. In particular, Figure \ref{HRdiagram} demonstrates locations of the selected stars on the Hertzsprung-Russell diagram; a more detailed analysis is presented in Appendix \ref{AppStellar}. The conclusions can be summarized as follows:

KIC 8093473 is located well above the main sequence, but is too faint to be a giant or subgiant. Most likely, it is an unresolved binary or multiple system consisting of several (from two to four) M dwarfs; the estimation of the number of components depends on the assumed age of the system (which is actually unknown). Determining the exact configuration of this system will require further observations. 

KIC 8454353 is located slightly above the main sequence. Most likely, it is an unresolved binary consisting of two more-or-less similar M dwarfs, with an age of $\gtrsim 100$ Myr.  

KIC 9048551 is located on the main sequence and can be identified as a single K dwarf, with an estimated age of $\sim 120-280$ Myr. 

Since about $30-40\%$ of K and M dwarfs in the solar neighborhood have been found to form binary or multiple systems \citep[e.g.,][]{Raghavan2010, Winters2019}, it is not surprising that two of three stars in our sample belong to this category, too. We note that, for an unresolved binary or multiple system, the parameters $T_{\mathrm{eff}}$ and $L$ in the \textit{Gaia} catalog (see Table \ref{stars}) represent respectively the average temperature and total luminosity of the system \citep[cf.][]{Andrae2018}. We have analyzed long-term \textit{Kepler} light curves of KIC 8093473 and KIC 8454353, and found no significant secondary rotation periods in both cases, which implies that either the components of these systems are tidally locked (for KIC 8093473, this favors a binary), or the secondary components are inactive. In Section \ref{BinarySection}, we demonstrate that an orbital interaction in the KIC 8093473 and KIC 8454353 systems is unlikely to affect the observed flares. All stars in our sample are relatively cool (K and M dwarfs) and rapidly rotating; i.e., they belong to the category of stars where superflares occur most frequently \citep[e.g.,][]{Davenport2016, VanDoorsselaere2017, Brasseur2019}.

\begin{figure*}
\centerline{\includegraphics{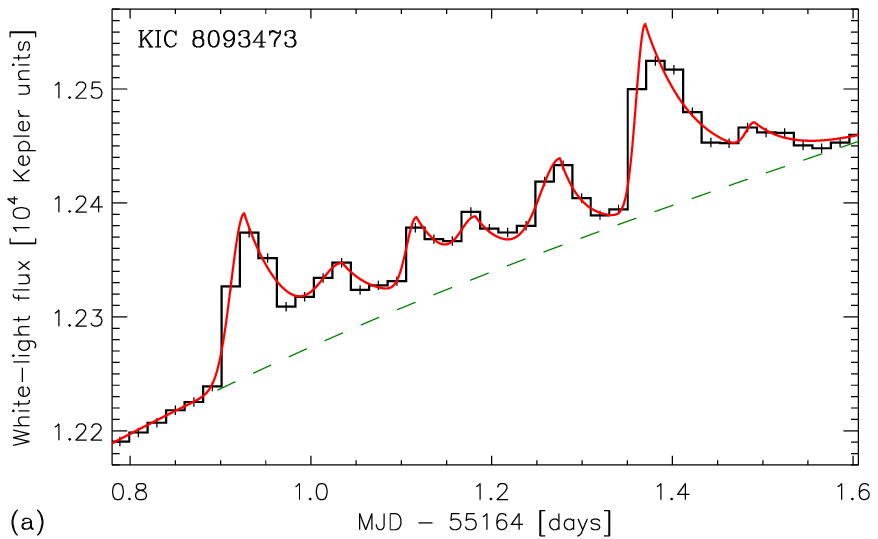}~
\includegraphics{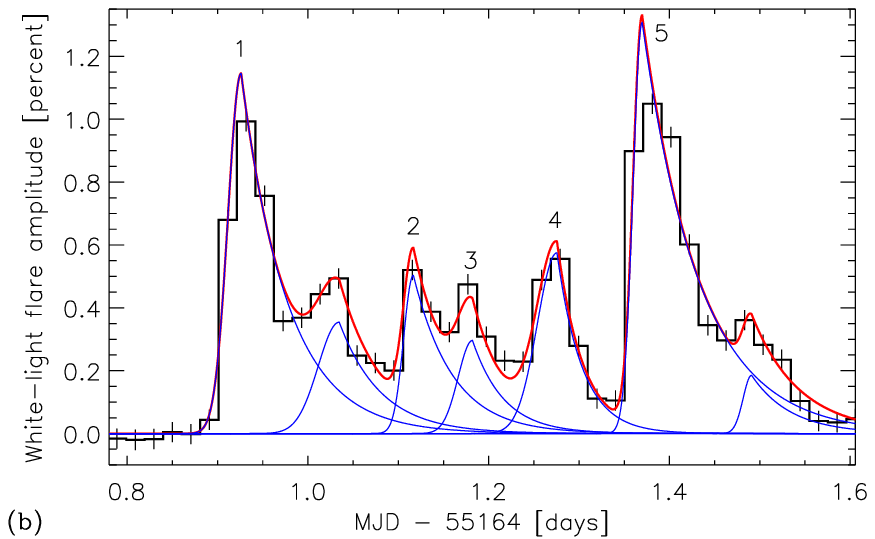}}
\centerline{\includegraphics{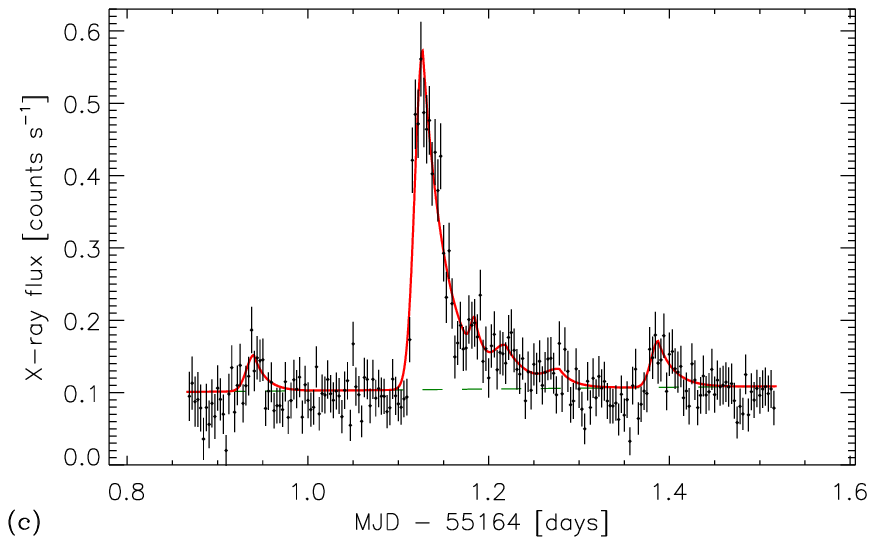}~
\includegraphics{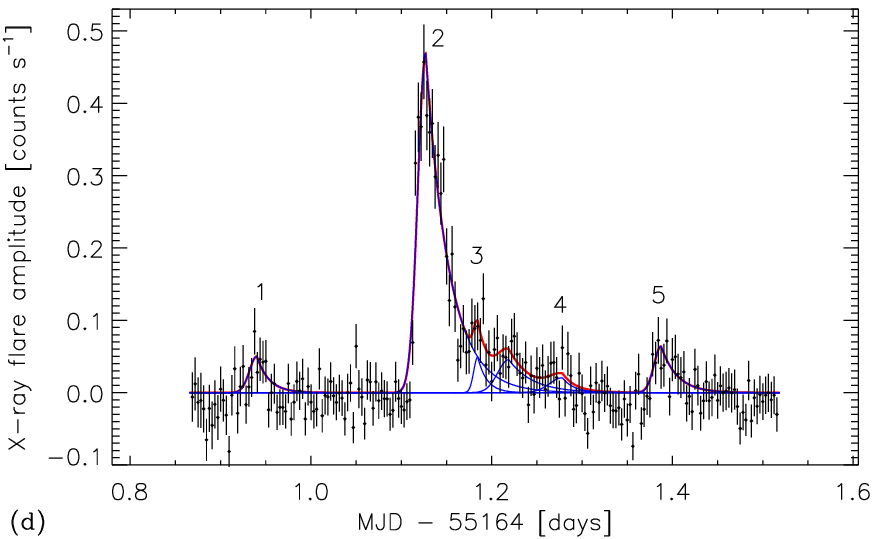}}
\caption{Light curves of KIC 8093473. (a) Optical (\textit{Kepler}) light curve. (b) Background-subtracted and normalized (by the average stellar flux) optical light curve. (c) X-ray (\textit{XMM-Newton} EPIC PN, $0.2-12$ keV) light curve. (d) Background-subtracted X-ray light curve. The error bars correspond to $1\sigma$ level. The fitted model light curve and the quiescent background light curve are shown by solid red and dashed green lines, respectively. Light curves of individual flaring components are shown by thin solid blue lines. The numbers near the peaks in panels (b) and (d) correspond to the flare numbers in Table \protect\ref{flares}.}
\label{LC8093473}
\end{figure*}

\begin{figure*}
\centerline{\includegraphics{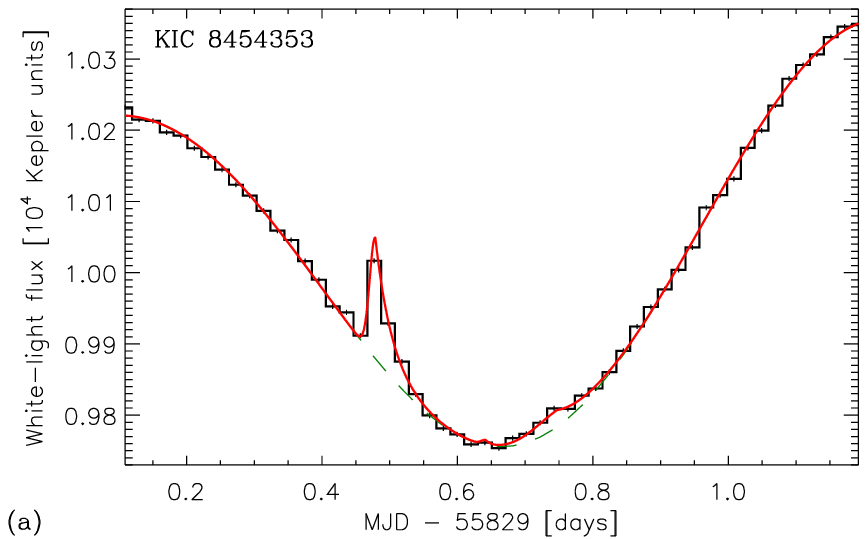}~
\includegraphics{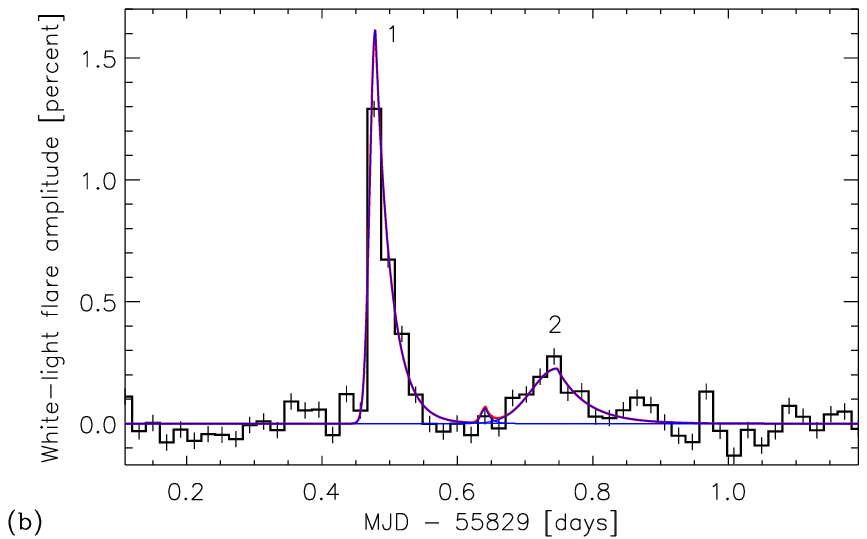}}
\centerline{\includegraphics{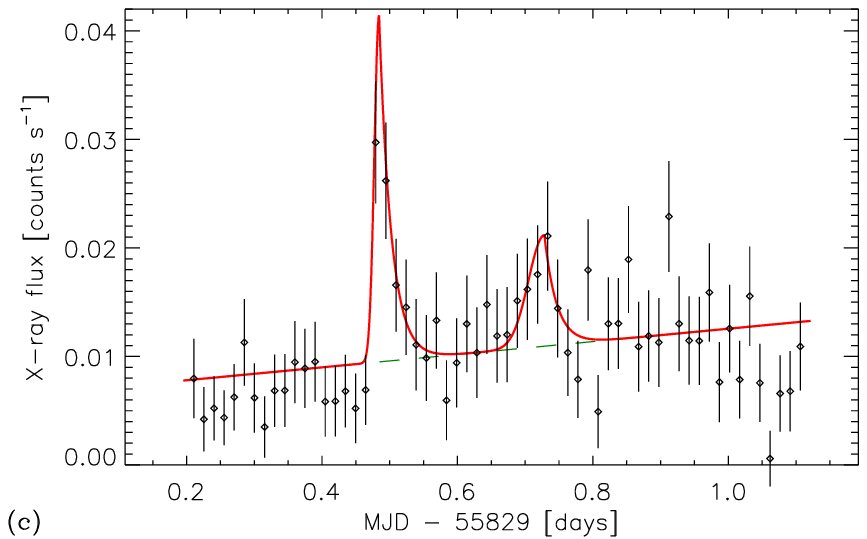}~
\includegraphics{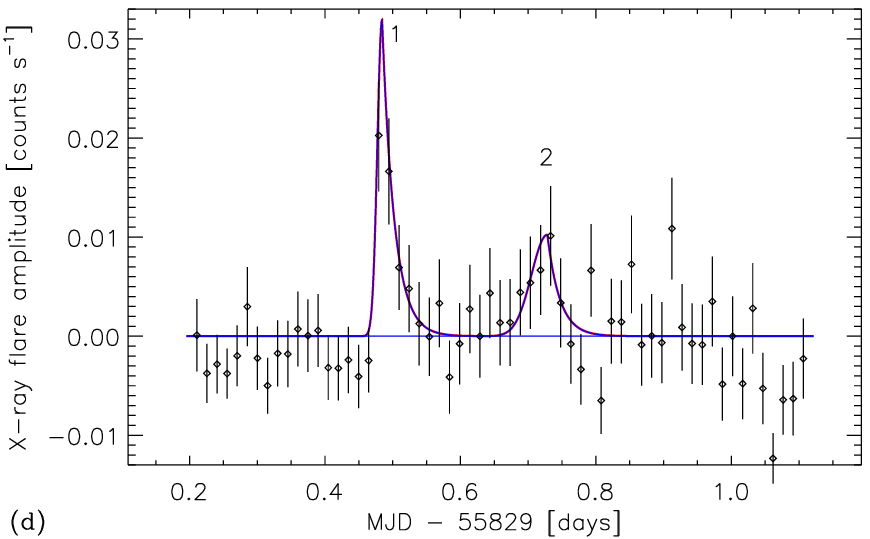}}
\caption{Same as in Figure \protect\ref{LC8093473}, for KIC 8454353.}
\label{LC8454353}
\end{figure*}

\begin{figure*}
\centerline{\includegraphics{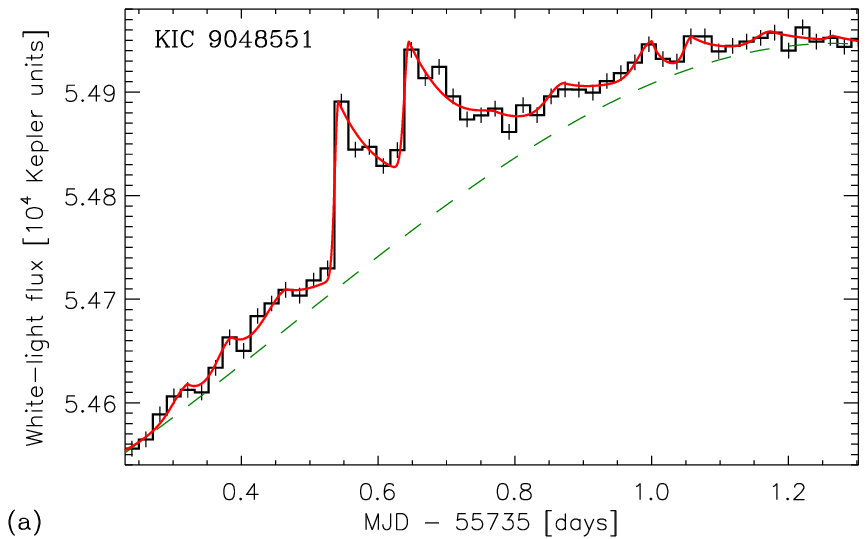}~
\includegraphics{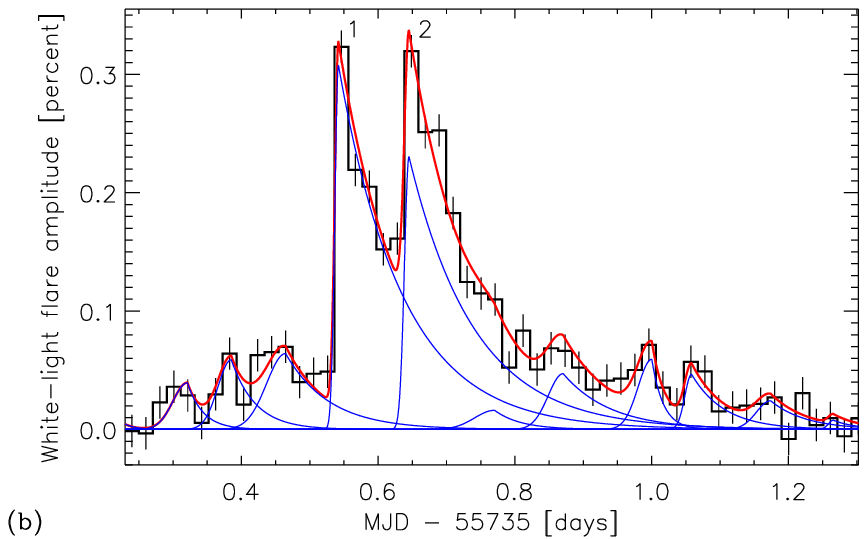}}
\centerline{\includegraphics{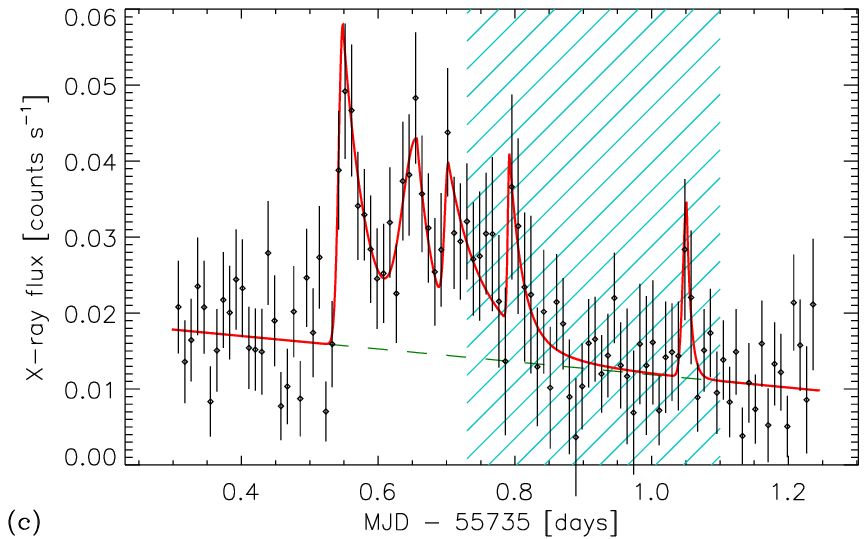}~
\includegraphics{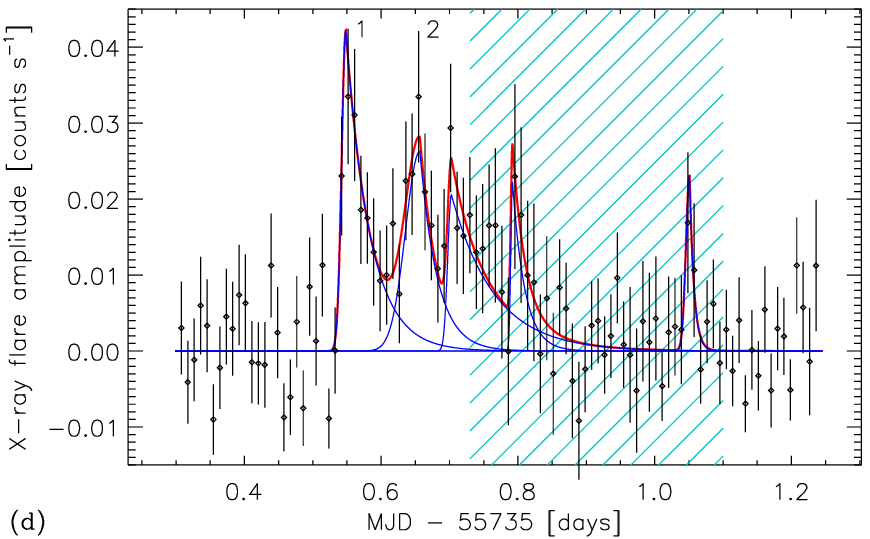}}
\caption{Same as in Figure \protect\ref{LC8093473}, for KIC 9048551. The shaded region in panels (c) and (d) represents the time interval when significant background fluctuations (of non-stellar origin) occurred.}
\label{LC9048551}
\end{figure*}

\subsection{Detected flares}
Figures \ref{LC8093473}--\ref{LC9048551} demonstrate the light curves of the selected stars vs.\ time in Modified Julian Day (MJD). For the X-rays, we consider here and below the data from the \textit{XMM-Newton} EPIC PN detector \citep{Struder2001} only, because it is the most sensitive one; the instrumental background is subtracted. The \textit{Kepler} light curves have the cadence of about 29.5 min; the bin sizes of the \textit{XMM-Newton} light curves vary from 270 to 1290 s for different objects, depending on the X-ray flux \citep{Rosen2016}. The gradual trends in the ``raw'' optical light curves (panels (a)) are caused by the rotational modulation (influence of starspots). The main features of the light curves are summarized below.

KIC 8093473 (Figure \ref{LC8093473}): there are two prominent white-light flares at around MJD 55164.93 and 55165.37, which are accompanied by weak but noticeable X-ray counterparts. The brightest X-ray flare occurs at around MJD 55165.13, with several weaker quasi-periodic peaks occurring in its decay phase; both the first X-ray flaring peak and the subsequent weaker peaks are accompanied by nearly simultaneous white-light brightenings. Such a behaviour is typical of large solar and stellar flares, with the multiple peaks either caused by the modulation of the flaring emissions by magnetohydrodynamic oscillations, or representing separate acts of magnetic reconnection (sub-flares) in the same active region \citep[see, e.g., the recent review by][]{Kupriyanova2020}. Notably, the flares at MJD 55164.93 and 55165.37 are more pronounced in the optical range than in the X-rays, while the flare at MJD 55165.13 is more pronounced in the X-rays than in the optical range; we discuss this peculiarity in Section \ref{WLXsection}.

KIC 8454353 (Figure \ref{LC8454353}): there is a prominent sharp flare at around MJD 55829.48, occurring simultaneously in the optical and X-ray ranges. Another weaker but slightly longer X-ray flare occurs at around MJD 55829.75; it is accompanied by a very faint but noticeable white-light counterpart.

KIC 9048551 (Figure \ref{LC9048551}): there are two prominent partially overlapping flares at around MJD 55735.54 and 55735.65, occurring nearly simultaneously in the optical and X-ray ranges (the X-ray flares are slightly shorter); several weaker peaks are visible at later times in both  spectral ranges. The shaded region in the X-ray plots represents the time interval when significant background X-ray fluctuations (with the flux comparable to that from the target star) occurred, which makes the data less reliable; for this reason, we do not analyze here the flares that occurred during the mentioned background event.


\subsection{Flare occurrence vs. the rotation phase}
There was no noticeable correlation between the flares and the stellar rotational phase: the flares occurred both around the local minimum (for KIC 8454353) and near the local maxima (for KIC 8093473 and KIC 9048551) of the long-term \textit{Kepler} light curves. Although the number of flares in our sample is rather small, the absence of a preferable rotational phase is consistent with the results of the large-scale statistical study by \citet{Doyle2018, Doyle2019, Doyle2020}. Most likely, similarly to the Sun, the considered stars possess multiple active regions (and hence multiple starspots), and flares do not necessarily occur in the largest of them.

\section{Methods}
\subsection{Fitting of the light curves}\label{LCfitting}
To analyze the identified stellar flares quantitatively, we best-fitted the observed light curves with model ones. The model that we used represents a flaring component superimposed on a variable quiescent background: $I(t)=I_{\mathrm{back}}(t)+I_{\mathrm{flare}}(t)$. Each flare (in either optical or X-ray range) was modeled as an asymmetric peak with a Gaussian rise phase and an exponential decay phase; i.e., the contribution of the flares had the form:

\begin{eqnarray}
I_{\mathrm{flare}}(t) & = & \sum\limits_{i=1}^NI_{\mathrm{flare}}^{(i)}(t)\nonumber\\
& = & \sum\limits_{i=1}^Ng_i\left\{\begin{array}{ll}
\displaystyle\exp\left[-\frac{(t-t_i^0)^2}{(\tau_i^{\mathrm{rise}})^2}\right], & t<t_i^0,\\[10pt]
\displaystyle\exp\left(-\frac{t-t_i^0}{\tau_i^{\mathrm{decay}}}\right), & t\ge t_i^0,
\end{array}\right.\label{FlareMod}
\end{eqnarray}
where $N$ is the number of flares revealed for each star. For the optical observations, the number $N$ was determined by the number of local maxima in the original light curves. To account for different time resolutions and select the events that revealed themselves in both spectral ranges, searching for flares in the X-ray light curves included an additional step: we firstly smoothed a light curve by filtering out the high-frequency component (with timescales shorter than 30 min, which corresponds to the \textit{Kepler} cadence), and then all local maxima exceeding an $1\sigma$ threshold in the residual signal were considered as potential flare candidates and their number determined the number $N$ in Equation (\ref{FlareMod}). This model implies that all significant peaks in a light curve (e.g., both the brightest X-ray flare on KIC 8093473 and the weaker peaks occurring in its decay phase, see Figure \ref{LC8093473}(c--d)) are considered as separate flares; this approach is consistent with the concept of ``complex flares'' used by \citet{Davenport2014}, and may also be attributed to a ``build-up and release'' scenario for flares in the solar and stellar coronae \citep{Hudson2020}.

In the optical range, the background variations are caused mainly by the stellar rotation. Therefore, for KIC 8093473 and KIC 9048551, we modeled the quiescent optical background with the following function:
\begin{eqnarray}
I_{\mathrm{back}}^{\mathrm{WL, 1}}(t) & = & A_1\sin\left(\frac{2\pi t}{P_{\mathrm{rot}}}+\varphi_1\right)\nonumber\\
&& +A_2\sin\left(\frac{4\pi t}{P_{\mathrm{rot}}}+\varphi_2\right)+C,\label{BackMod1}
\end{eqnarray}
where $P_{\mathrm{rot}}$ is the rotation period determined from earlier studies (see Table \ref{stars}). Equation (\ref{BackMod1}) represents a truncated Fourier expansion, including the first and second harmonics of the rotation frequency. For KIC 8454353, better agreement with the observations was achieved using a different background model function:
\begin{equation}\label{BackMod2}
I_{\mathrm{back}}^{\mathrm{WL, 2}}(t)=(A_0t^{A_1}+A_2)\sin\left(\frac{2\pi t}{P_{\mathrm{rot}}}+\varphi\right)+C.
\end{equation}
In this case, the gradual variation (increase) of the modulation depth with time could be attributed to the evolution of the stellar active regions (starspots). The \textit{Kepler} light curves were truncated to the time windows slightly extending those of the \textit{XMM-Newton} observations, as shown in Figures 1--3. For the X-ray emission, we adopted a simple linear model of the quiescent background: $I_{\mathrm{back}}^{\mathrm{XMM}}(t)=At+B$.

The model light curves were fitted to the observations using the Bayesian inference and Markov chain Monte Carlo (MCMC) sampling \citep[see, e.g.,][and references therein]{Gregory2010}. In this work, we used the MCMC sampling implementation by \citet{Pascoe2017} and \citet{Anfinogentov2021}. All the model parameters, except the rotation period $P_{\mathrm{rot}}$ and the number of flares $N$ (that was determined by the number of local maxima prior to the fitting procedure, as described above), were considered as free parameters. As an initial guess for the flare parameters in Equation (\ref{FlareMod}), we used positions of the local maxima (for $t_i^0$), the corresponding flare fluxes (for $g_i$), and the time intervals between the neighbouring apparent flare peaks (for $\tau_i^{\mathrm{rise}}$ and $\tau_i^{\mathrm{decay}}$). Then, for the X-ray emission, the MCMC fitting procedure used the original (non-smoothed) light curves.

To cope with the original long-cadence observations and sample the flare shapes properly, we used the method of supersampling, i.e., the model light curves were initially calculated on a fine time grid with 10 s cadence. After that, the model fine-resolution light curves were binned over the instrument exposure intervals, and then compared with the observations to evaluate the residuals and the likelihood function through the MCMC sampling algorithm.

\begin{deluxetable*}{cCCCCCCCCCCCC}
\renewcommand{\tabcolsep}{2.5pt}
\tablewidth{0pt}
\tablecaption{Parameters of the flares that occurred simultaneously in the white-light (WL) and X-ray (X) ranges: peak times ($t_0$), rise times ($\tau_{\mathrm{rise}}$), decay times ($\tau_{\mathrm{decay}}$), peak luminosities ($L_{\max}$), emitted energies ($E_{\mathrm{flare}}$), and peak equivalent \textit{GOES} X-ray fluxes ($I_{\max}^{\mathit{GOES}}$).\label{flares}}
\tablehead{\colhead{No.} & \colhead{$t_0^{\mathrm{WL}}$,} & \colhead{$\tau_{\mathrm{rise}}^{\mathrm{WL}}$,} & \colhead{$\tau_{\mathrm{decay}}^{\mathrm{WL}}$,} & \colhead{$L_{\max}^{\mathrm{WL}}$,} & \colhead{$E^{\mathrm{WL}}_{\mathrm{flare}}$,} & \colhead{$t_0^{\mathrm{X}}$,} & \colhead{$\tau_{\mathrm{rise}}^{\mathrm{X}}$,} & \colhead{$\tau_{\mathrm{decay}}^{\mathrm{X}}$,} & \colhead{$L_{\max}^{\mathrm{X}}$,} & \colhead{$I_{\max}^{\mathit{GOES}}$,} & \colhead{$E^{\mathrm{X}}_{\mathrm{flare}}$,}\\[-6pt]
\colhead{} & \colhead{days} & \colhead{min} & \colhead{min} & \colhead{$10^{28}$ erg~$\textrm{s}^{-1}$} & \colhead{$10^{32}$ erg} & \colhead{days} & \colhead{min} & \colhead{min} & \colhead{$10^{28}$ erg~$\textrm{s}^{-1}$} & \colhead{$10^{-2}$ W~$\textrm{m}^{-2}$} & \colhead{$10^{32}$ erg}}
\startdata
\multicolumn{12}{c}{KIC 8093473}\\
1 & 0.926_{0.917}^{0.935} &  29.6_{ 11.2}^{ 45.0} &  73.3_{ 46.1}^{104.0} &  83.2_{ 56.8}^{202.5} &  46.0_{ 33.4}^{111.1} & 0.940_{0.940}^{0.942} & 16.7_{ 9.1}^{37.6} &  19.3_{ 18.4}^{ 29.7} &  63.2_{ 21.7}^{ 89.5} &  29.5_{ 10.3}^{ 41.3} &  10.6_{  6.5}^{ 17.7}\\
2 & 1.117_{1.107}^{1.134} &  21.9_{  7.3}^{ 57.2} &  62.7_{ 16.6}^{ 65.6} &  36.6_{ 15.0}^{ 97.0} &  15.1_{  8.6}^{ 38.4} & 1.127_{1.127}^{1.129} & 17.8_{16.9}^{21.1} &  36.8_{ 31.0}^{ 39.7} & 434.0_{357.8}^{523.7} & 147.1_{127.1}^{170.7} & 135.1_{114.8}^{157.8}\\
3 & 1.182_{1.166}^{1.199} &  32.4_{ 13.2}^{100.7} &  50.2_{ 13.9}^{108.5} &  21.5_{  7.9}^{ 44.1} &  12.9_{  6.8}^{ 34.7} & 1.185_{1.183}^{1.186} &  9.0_{ 6.9}^{23.1} &  14.6_{ 14.4}^{ 23.3} &  45.4_{ 37.8}^{ 93.3} &  15.4_{ 13.4}^{ 30.4} &   8.9_{  6.7}^{ 13.3}\\
4 & 1.276_{1.265}^{1.293} &  44.2_{ 19.8}^{ 96.7} &  40.2_{ 10.1}^{ 79.9} &  41.8_{ 22.0}^{ 92.7} &  16.6_{ 11.3}^{ 46.5} & 1.278_{1.277}^{1.279} & 29.3_{ 9.4}^{37.6} &  18.5_{ 16.4}^{ 26.6} &  19.0_{ 10.3}^{ 36.3} &   6.4_{  3.6}^{ 11.8} &   4.7_{  3.2}^{  8.4}\\
5 & 1.370_{1.364}^{1.379} &  20.4_{ 10.9}^{ 38.3} &  91.4_{ 68.9}^{108.7} &  95.0_{ 66.0}^{217.8} &  61.6_{ 45.1}^{138.1} & 1.388_{1.386}^{1.389} & 16.0_{11.8}^{31.7} &  25.5_{ 24.6}^{ 39.8} &  68.2_{ 25.4}^{ 92.5} &  27.3_{ 10.4}^{ 36.3} &  15.1_{  9.5}^{ 21.9}\\[3pt]
\multicolumn{12}{c}{KIC 8454353}\\
1 & 0.479_{0.477}^{0.482} &  16.9_{ 16.7}^{ 22.5} &  33.2_{ 23.6}^{ 41.4} &  61.2_{ 46.0}^{146.8} &  17.6_{ 13.9}^{ 40.4} & 0.484_{0.477}^{0.498} & 14.2_{12.4}^{41.5} &  25.9_{ 26.3}^{ 70.5} &  43.1_{ 18.5}^{ 56.3} &  16.2_{  5.7}^{ 23.5} &  10.4_{  6.6}^{ 16.3}\\
2 & 0.747_{0.736}^{0.750} &  77.0_{ 22.8}^{ 98.9} &  63.0_{ 15.0}^{ 72.6} &   8.6_{  4.8}^{ 32.6} &   6.2_{  3.7}^{ 18.1} & 0.728_{0.661}^{0.742} & 47.7_{13.2}^{64.7} &  26.6_{ 26.9}^{158.6} &   9.5_{  8.3}^{ 18.0} &   1.1_{  0.8}^{  2.3} &   5.3_{  3.9}^{  9.3}\\[3pt]
\multicolumn{12}{c}{KIC 9048551}\\
1 & 0.542_{0.541}^{0.547} &  10.7_{ 10.6}^{ 20.1} & 135.2_{ 74.9}^{139.2} &  25.6_{ 20.0}^{ 52.1} &  18.2_{ 14.0}^{ 36.5} & 0.548_{0.543}^{0.559} & 12.3_{ 5.3}^{27.3} &  50.2_{ 29.1}^{ 66.6} &  13.0_{  8.5}^{ 15.9} &   1.1_{  0.6}^{  1.4} &   4.5_{  3.6}^{  5.4}\\
2 & 0.645_{0.643}^{0.655} &  13.6_{ 13.3}^{ 33.8} & 138.1_{ 85.7}^{164.9} &  19.2_{ 15.2}^{ 41.9} &  17.3_{ 13.1}^{ 35.4} & 0.657_{0.640}^{0.665} & 42.6_{14.0}^{56.2} &  36.7_{ 16.2}^{118.9} &   8.1_{  5.8}^{ 11.8} &   0.7_{  0.4}^{  1.0} &   4.0_{  3.0}^{  5.9}\\
\enddata
\tablecomments{The flare peak times $t_0^{\mathrm{WL}}$ and $t_0^{\mathrm{X}}$ for KIC 8093473, KIC 8454353, and KIC 9048551 are relative to MJD 55164, MJD 55829, and MJD 55735, respectively.}
\end{deluxetable*}

The MCMC fitting provided the most probable values of the model parameters, as well as robust estimations of their confidence intervals. Parameters of the flares that occurred simultaneously in the optical and X-ray ranges are presented in Table \ref{flares}; all significant peaks in the light curves revealed by the above-described analysis are listed in Table \ref{allflares} in Appendix \ref{AppData}. In Figures \ref{LC8093473}--\ref{LC9048551}, panels (a) and (c), the best-fitted model light curves and the quiescent background components are overplotted on the observed data, while panels (b) and (d) demonstrate the flaring (background-subtracted) components.

\subsection{White-light flare parameters}\label{WLparameters}
We estimated the white-light flare luminosities and energies following the approach of \citet{Shibayama2013} and \citet{Namekata2017}. Namely, we assumed that the spectrum of a white-light flare can be described by a blackbody radiation with a temperature of $T_{\mathrm{flare}}$, the star itself is a blackbody source with a temperature of $T_{\mathrm{star}}$, and the normalized (by the average stellar flux) flare amplitude in the light curve is proportional to a fraction of the stellar disk covered by the white-light flare ribbons, with account for the different spectral shapes of the quiescent and flaring emissions as well as for the instrumental bandpass. This gives
\begin{equation}\label{IWL}
\frac{I_{\mathrm{flare}}^{\mathrm{WL}}(t)}{\left<I_{\mathrm{star}}^{\mathrm{WL}}\right>}=\frac{A_{\mathrm{flare}}(t)}{\pi R_{\mathrm{star}}^2}\frac{\int R_{\lambda}(\lambda)B_{\lambda}(\lambda, T_{\mathrm{flare}})\,\mathrm{d}\lambda}{\int R_{\lambda}(\lambda)B_{\lambda}(\lambda, T_{\mathrm{star}})\,\mathrm{d}\lambda},
\end{equation} 
where $I_{\mathrm{flare}}^{\mathrm{WL}}(t)$ is the background-subtracted \textit{Kepler} flare light curve, $\left<I_{\mathrm{star}}^{\mathrm{WL}}\right>$ is the average \textit{Kepler} stellar flux, $A_{\mathrm{flare}}(t)$ is the visible (projected) area of the flare ribbons, $R_{\mathrm{star}}$ is the stellar radius, $\lambda$ is the wavelength, $B_{\lambda}(\lambda, T)$ is the Planck function, and $R_{\lambda}(\lambda)$ is the \textit{Kepler} response function \citep[in the $350-950$ nm spectral band;][]{vanCleve2016}. As a result, the bolometric luminosity of a white-light flare $L_{\mathrm{flare}}^{\mathrm{WL}}(t)$ can be expressed in the form
\begin{eqnarray}
\frac{L_{\mathrm{flare}}^{\mathrm{WL}}(t)}{\left<L_{\mathrm{star}}^{\mathrm{WL}}\right>} & = & \frac{1}{4}\frac{I_{\mathrm{flare}}^{\mathrm{WL}}(t)}{\left<I_{\mathrm{star}}^{\mathrm{WL}}\right>}\nonumber\\
&& \times\frac{T^4_{\mathrm{flare}}}{T^4_{\mathrm{star}}}\frac{\int R_{\lambda}(\lambda)B_{\lambda}(\lambda, T_{\mathrm{star}})\,\mathrm{d}\lambda}{\int R_{\lambda}(\lambda)B_{\lambda}(\lambda, T_{\mathrm{flare}})\,\mathrm{d}\lambda},\label{LWLt}
\end{eqnarray}
where $\left<L_{\mathrm{star}}^{\mathrm{WL}}\right>$ is the average bolometric stellar luminosity; the factor $1/4$ arises because we consider the total stellar luminosity (for a spherical source), while the sources of the flaring white-light emission look like nearly flat patches on the stellar surface. For an unresolved binary or multiple system, the temperature $T_{\mathrm{star}}$ in Equations (\ref{IWL}--\ref{LWLt}) represents an average effective temperature of the system components, and the parameters $\pi R_{\mathrm{star}}^2$ and $\left<L_{\mathrm{star}}^{\mathrm{WL}}\right>$ should be replaced by the total visible area of the stellar disks ($\pi R_{\mathrm{A}}^2+\pi R_{\mathrm{B}}^2+\ldots$) and the total luminosity of the system ($\left<L_{\mathrm{A}}^{\mathrm{WL}}+L_{\mathrm{B}}^{\mathrm{WL}}+\ldots\right>$), respectively. We used the stellar temperature and luminosity values determined by \textit{Gaia} (see Table \ref{stars}), which can be applied to both single stars and unresolved systems, as described above; the average \textit{Kepler} fluxes were determined by averaging the light curves over the entire respective quarters. Following \citet{Maehara2012, Shibayama2013, Namekata2017}, we adopted $T_{\mathrm{flare}}=10\,000$ K as the typical effective temperature of the white-light flares, with possible variations in the range of $9000-14\,000$ K \citep{Kowalski2016}; this uncertainty in the flare temperature is responsible for about a half of the uncertainties in the estimated white-light flare energies and luminosities.

The total radiated flare energy (in either spectral band) is an integral of its luminosity:
\begin{equation}\label{Eflare}
E_{\mathrm{flare}}^{\textrm{WL,X}}=\int L_{\mathrm{flare}}^{\textrm{WL,X}}(t)\,\mathrm{d}t.
\end{equation}
To estimate the energies and peak luminosities of the individual flares, we used the fitted model light curves, i.e., $i$th components in Equation (\ref{FlareMod}). The most probable values and confidence intervals for the integrals under the flare light curves were determined from the MCMC fitting procedure.

\subsection{X-ray flare parameters}
The X-ray luminosity of a stellar source (for the flaring or/and quiescent emissions, see below) can be estimated as
\begin{equation}\label{LX}
L^{\mathrm{X}}=2\pi d^2\int\limits_{E_{\min}}^{E_{\max}}F(E_{\mathrm{ph}})E_{\mathrm{ph}}\,\mathrm{d}E_{\mathrm{ph}},
\end{equation}
where $d$ is the distance to the star, $E_{\mathrm{ph}}$ is the photon energy, $F(E_{\mathrm{ph}})$ is the model X-ray spectral flux density, and $E_{\min}\le E_{\mathrm{ph}}\le E_{\max}$ is the considered energy range ($0.2-12$ keV in this work). We note that both the model X-ray spectrum $F(E_{\mathrm{ph}})$ and the resulting luminosity $L^{\mathrm{X}}$ in Equation (\ref{LX}) are being averaged over a time interval where the spectral fitting is performed, i.e., $L^{\mathrm{X}}\equiv\left<L^{\mathrm{X}}\right>$.

As said above, we used the \textit{XMM-Newton} EPIC PN detector data. We extracted the time-resolved spectral data for the selected stars from the \textit{XMM-Newton} science archive\footnote{\url{https://www.cosmos.esa.int/web/xmm-newton/xsa}} using the same source and background regions and ``good'' time intervals (presented in the 3XMM-DR5 database) that were used to produce the light curves. Then we analyzed the X-ray spectra in selected time intervals (see below) using the OSPEX package \citep{Tolbert2020}; the spectra were fitted with a single-temperature optically thin thermal model (\textsc{vth}), with account for interstellar absorption \citep[computed using the model of][]{Morrison_1983}. Following \citet{Guarcello2019a, Pizzocaro2019}, etc., we estimated the absorption column density for each star as $N_{\mathrm{H}}=1.79\times 10^{21}A_V$ $\textrm{cm}^2$ $\textrm{mag}^{-1}$, where $A_V$ is the known optical extinction (see Table \ref{stars}). We set the abundances of heavy elements to $0.2$ of the solar ones---a typical value for the coronae of active stars \citep[e.g.,][]{Robrade2005, Pandey2012}. Considering the absorption column and/or abundances as free parameters has not significantly affected the obtained results.

Subtraction of the quiescent background is an important part of spectral analysis of flaring X-ray emission. However, for the considered observations, it was not possible to obtain reliable pre- or postflare background spectra. Instead, we assumed that the spectra of the flaring and quiescent emission components have similar shapes: e.g., for the optically thin thermal emission model, the plasma temperature remains nearly constant throughout time, and only the emission measure varies. This assumption can be justified for active K-M stars, where the quiescent X-ray emission is partially produced by the hot coronae (with the temperatures comparable to the temperatures during flares), and partially consists of multiple unresolved weaker flares \citep{Gudel2004}. In this case, the time-dependent X-ray flare luminosity $L_{\mathrm{flare}}^{\mathrm{X}}(t)$ can be estimated as
\begin{equation}\label{LXt}
\frac{L_{\mathrm{flare}}^{\mathrm{X}}(t)}{\left<L^{\mathrm{X}}_{\mathrm{total}}\right>}\simeq\frac{I_{\mathrm{flare}}^{\mathrm{X}}(t)}{\left<I^{\mathrm{X}}_{\mathrm{total}}\right>},
\end{equation}
where $I_{\mathrm{flare}}^{\mathrm{X}}(t)$ is the background-subtracted \textit{XMM-Newton} flare light curve, $\left<I^{\mathrm{X}}_{\mathrm{total}}\right>$ is the average total (i.e., including both the flares and the quiescent background) \textit{XMM-Newton} flux in the selected time interval, and $\left<L^{\mathrm{X}}_{\mathrm{total}}\right>$ is the average total X-ray luminosity in the selected time interval (Equation (\ref{LX})). This approach is certainly an approximate one; however, as demonstrated, e.g., by \citet{Flaccomio2018}, the inaccuracy introduced due to neglecting the plasma temperature variations during flares is small with respect to other sources of uncertainties, such as the measurement and spectral fitting uncertainties.

We performed spectral fitting of the \textit{XMM-Newton} spectra in the ``flaring'' time intervals, each selected to cover either one flare or several overlapping flares; these time intervals and the resulting best-fit spectral parameters are presented in Table \ref{xrayfits} in Appendix \ref{AppData}. We estimated the radiated energies and peak luminosities of the individual X-ray flares in the same way as for the white-light flares, i.e., using Equation (\ref{Eflare}) and the fitted model light curves.

We estimated also the equivalent \textit{GOES} fluxes---i.e., the X-ray flare fluxes as if observed by the \textit{GOES} satellite from a distance of 1 AU. They are given by the expression $I^{\mathit{GOES}}_{\textrm{flare}}(t)=L^{\mathit{GOES}}_{\textrm{flare}}(t)/(2\pi\mathcal{R}^2)$, where $\mathcal{R}=1$ AU, and the luminosities $L^{\mathit{GOES}}_{\textrm{flare}}(t)$ are computed as described above, but for the \textit{GOES} energy range ($1.55-12.4$ keV or $1-8$ {\AA}). Since the X-ray fluxes from the stellar flares were measured reliably only at the energies of up to a few keV, the estimations of the equivalent \textit{GOES} fluxes are based largely on extrapolation.

\section{Results and discussion}
Table \ref{flares} summarizes the parameters of the flares that occurred simultaneously in the optical and X-ray ranges. In total, we identified nine such events, including partially overlapping ones. The total radiated energies of the flares (in both spectral ranges) varied from $\sim 1.2\times 10^{33}$ to $\sim 1.5\times 10^{34}$ erg, which puts them into the category of superflares \citep{Maehara2012}. The peak X-ray fluxes were equivalent to the \textit{GOES} classes from $\sim\textrm{X}70$ to $\sim\textrm{X}14\,700$ (for comparison, the largest observed solar X-ray flare was of X28 class). Scatter plots demonstrating mutual correlations between various flare parameters are shown in Figure \ref{AllCorrelations} in Appendix \ref{AppData}; below, we examine some of these correlations in detail.

\begin{figure*}
\centerline{\includegraphics{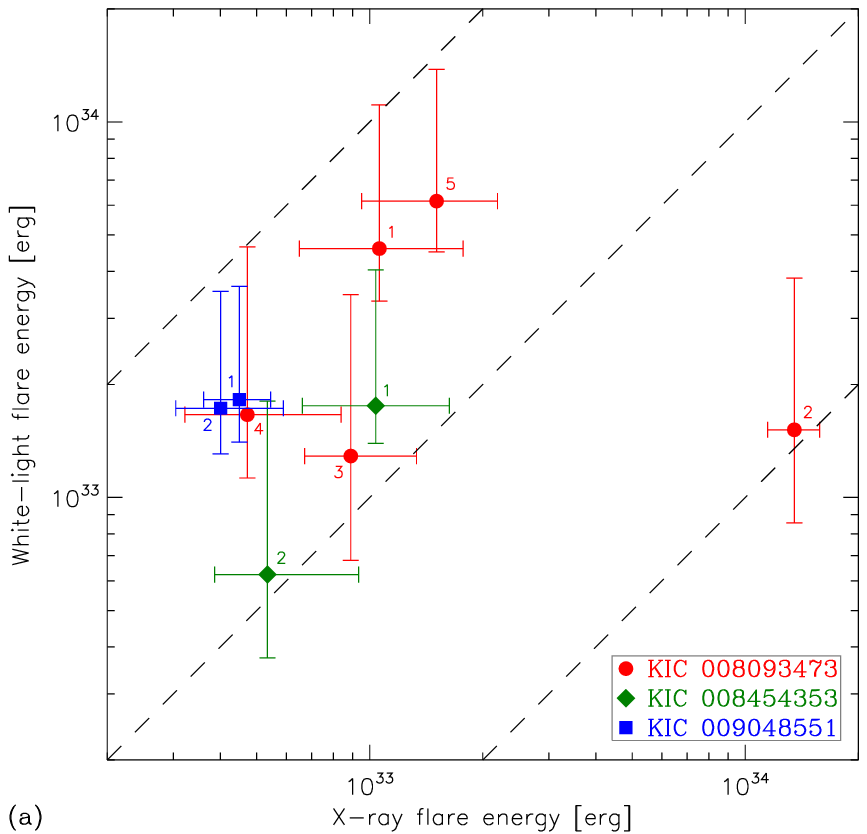}~
\includegraphics{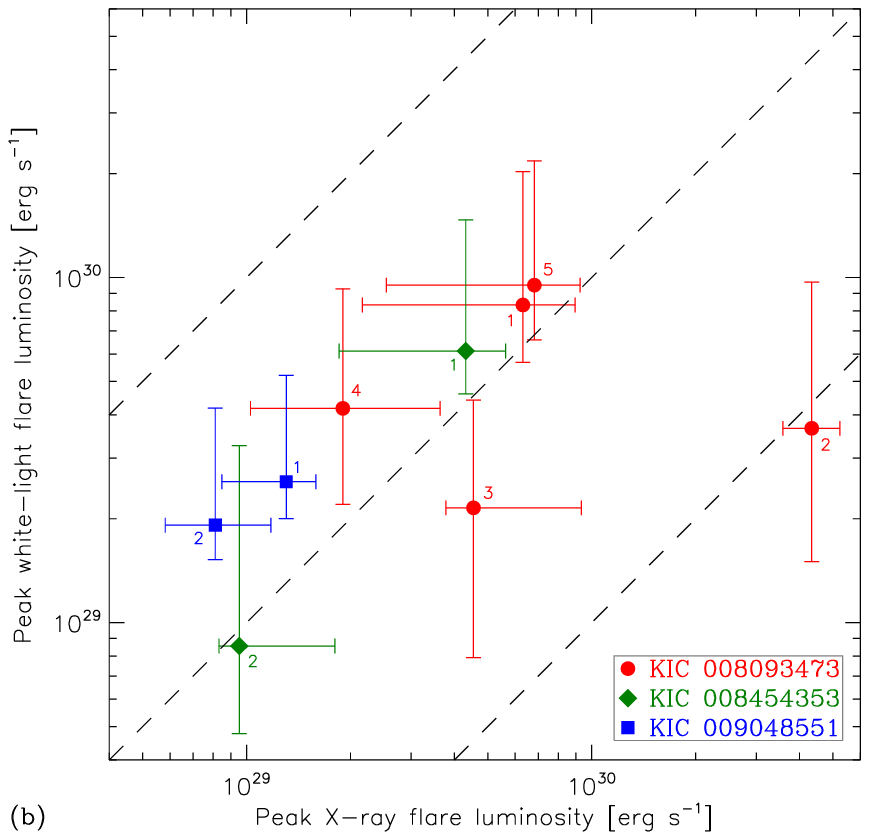}}
\caption{Comparison of the flare energies (a) and peak luminosities (b) in the X-ray and white-light ranges. The flare numbers correspond to those in Table \protect\ref{flares}. The error bars correspond to $1\sigma$ level. The three diagonal dashed lines represent the 0.1, 1, and 10 ratios between the plotted values.}
\label{Energetics}
\end{figure*}

\begin{figure*}
\centerline{\includegraphics{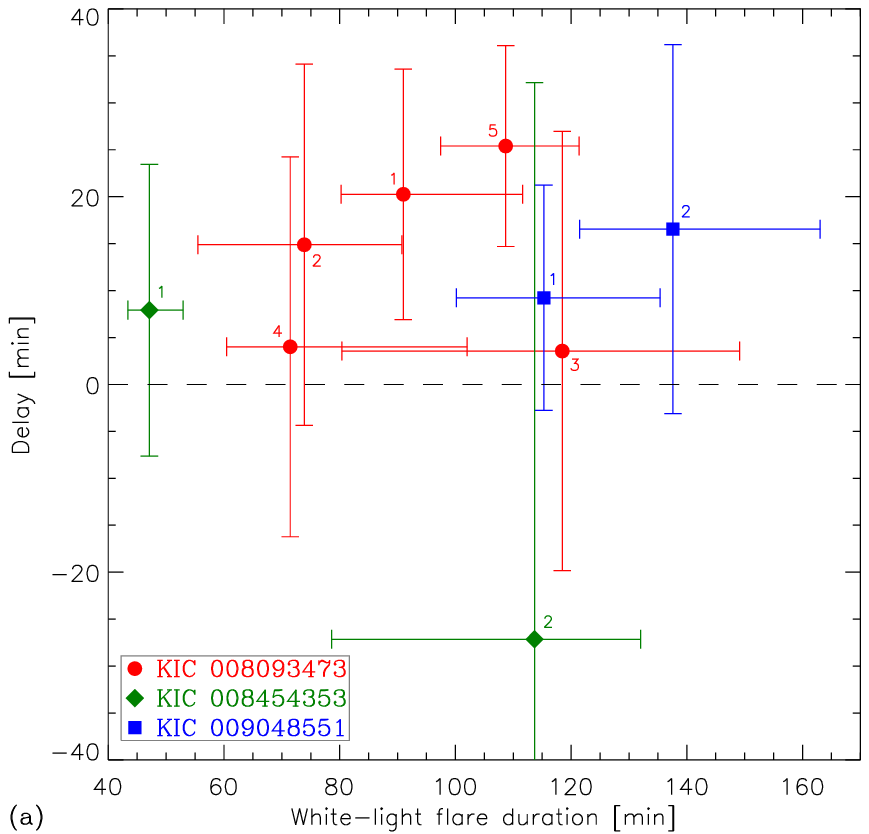}~
\includegraphics{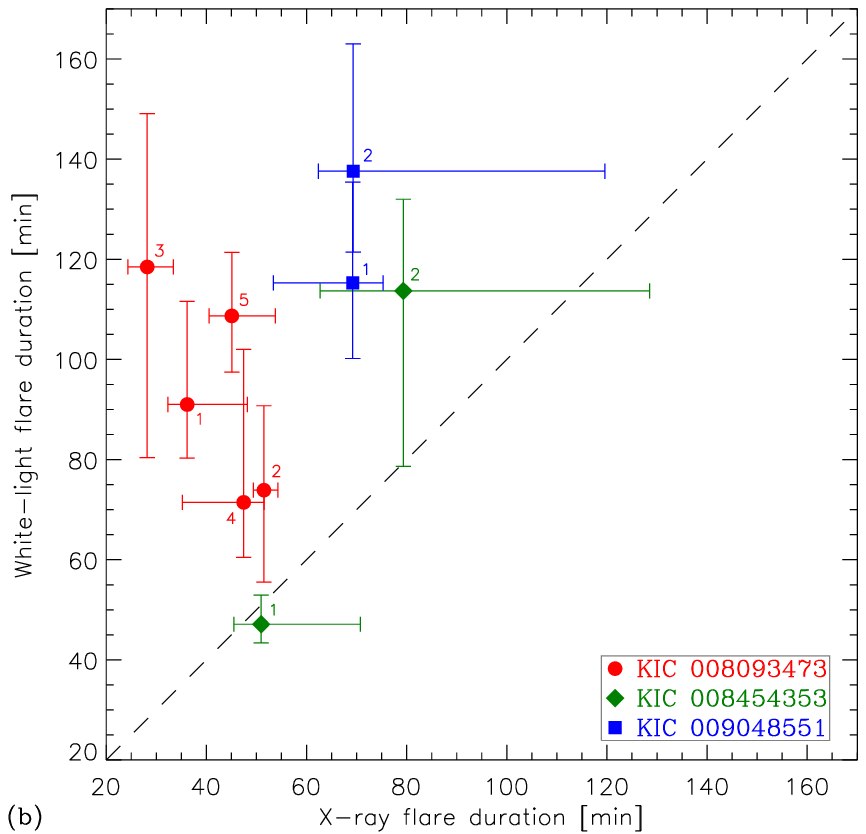}}
\caption{(a) Delays of the X-ray flares with respect to the corresponding optical flares $t_0^{\mathrm{X}}-t_0^{\mathrm{WL}}$ vs. the white-light flare durations. (b) Comparison of the flare durations in the X-ray and white-light ranges. The flare numbers correspond to those in Table \protect\ref{flares}. The error bars correspond to $1\sigma$ level.}
\label{Times}
\end{figure*}

\subsection{Comparison of the white-light and X-ray flare parameters}\label{WLXsection}
Figure \ref{Energetics} demonstrates scatter plots of the radiated flare energies and peak flare luminosities in the optical and X-ray ranges. Most of the analyzed flares released more energy in the optical range than in the X-ray one: $E^{\mathrm{WL}}_{\mathrm{flare}}/E^{\mathrm{X}}_{\mathrm{flare}}\sim 3-4$ in five flares, and $E^{\mathrm{WL}}_{\mathrm{flare}}/E^{\mathrm{X}}_{\mathrm{flare}}\sim 1.5$ in three flares. Similarly, the peak white-light flare luminosities were usually higher than or comparable to the X-ray ones: $L^{\mathrm{WL}}_{\max}/L^{\mathrm{X}}_{\max}\sim 1-2$ in seven flares. This energy partition is typical of the solar flares, where the white-light continuum emission is responsible, on average, for about 70\% of the total radiated flare energies \citep{Kretzschmar2011}. Similar relations between the optical and X-ray emissions in stellar flares were reported, e.g., by \citet{Fuhrmeister2011, Flaccomio2018, Guarcello2019a, Guarcello2019b, Schmitt2019}\footnote{We note that all mentioned estimations for the solar and stellar flares refer only to the energy released in the form of electromagnetic emission, and do not include other possible sinks of the flare energy such as escaping energetic particles and kinetic energy of coronal mass ejections.}.

A prominent outlier is the powerful flare \#2 on KIC 8093473, where the X-ray emission strongly dominated: $E^{\mathrm{WL}}_{\mathrm{flare}}/E^{\mathrm{X}}_{\mathrm{flare}}\simeq 1/9$ and $L^{\mathrm{WL}}_{\max}/L^{\mathrm{X}}_{\max}\simeq 1/12$. This difference from the other flares looks even more intriguing because the flare \#2 together with subsequent weaker flares \#3 and \#4 were likely parts of one long complex event (see Figure \ref{LC8093473}); however, the flares \#3 and \#4 demonstrated more typical relations between the optical and X-ray emissions (with $E^{\mathrm{WL}}_{\mathrm{flare}}\gtrsim E^{\mathrm{X}}_{\mathrm{flare}}$ and $L^{\mathrm{WL}}_{\max}\sim L^{\mathrm{X}}_{\max}$). In fact, in the sequence of flares \#2--4 on KIC 8093473, the $E^{\mathrm{WL}}_{\mathrm{flare}}/E^{\mathrm{X}}_{\mathrm{flare}}$ and $L^{\mathrm{WL}}_{\max}/L^{\mathrm{X}}_{\max}$ ratios tended to increase with time. We propose two explanations of this phenomenon:

a) ``Soft electron spectrum''. The white-light emission in the flare \#2 on KIC 8093473 could be relatively weak, if the accelerated electrons in this flare, despite of a large total energy flux, had a relatively soft spectrum; therefore, these electrons heated a large amount of plasma in the lower corona (which produced the soft X-rays), but were unable to penetrate into the deeper layers of the stellar atmosphere where the white-light emission is produced. According to the gas-dynamic simulations of \citet{Katsova1980}, the optical continuum emission should be negligible for the electron beams with the spectral indices of $\delta\gtrsim 4.5$; due to the limited spectral coverage of the observations, we cannot estimate the parameters of nonthermal energetic particles in the analyzed flares independently. If the variations of the $E^{\mathrm{WL}}_{\mathrm{flare}}/E^{\mathrm{X}}_{\mathrm{flare}}$ ratio from flare to flare were indeed caused by the mentioned effect, the spectra of energetic electrons in the long complex event including the flares \#2-4 on KIC 8093473 should have become increasingly harder with time---a behaviour (``soft-hard-harder'') that has also been observed in some solar flares \citep[see, e.g.,][]{Fletcher2011}.

b) ``Limb flare''. The observed white-light emission depends on the flare location on the stellar disk: the area $A_{\mathrm{flare}}$ in Equation (\ref{IWL}) is the projected area of the optically thick emission source, which decreases with the distance from the disk center and approaches zero for the flares at the limb. In contrast, the optically thin soft X-ray emission is not affected by the source location (unless the emitting volume is partially occulted). Therefore, the observed ratios $L^{\mathrm{WL}}_{\mathrm{flare}}/L^{\mathrm{X}}_{\mathrm{flare}}$ and $E^{\mathrm{WL}}_{\mathrm{flare}}/E^{\mathrm{X}}_{\mathrm{flare}}$ are expected to decrease significantly for the flares near the limb; this effect is confirmed by observations of solar flares, where the flare location is known \citep[e.g.,][]{Woods2006}. Thus the flare \#2 on KIC 8093473 could be an example of flare that occurred near the stellar limb, so that only a small fraction ($\sim 0.03$) of the total white-light emission was observed. In this case, the subsequent flares \#3 and \#4 should have occurred at different locations, approaching the stellar disk center with time. We note that similar (albeit smaller) shifts of the flare ribbons have been observed in solar flares: as demonstrated, e.g., by \citet{Grigis2005, Zimovets2009, Zimovets2013, Kuznetsov2016}, the energy release sites responsible for different flaring peaks in long complex events (at least, in some of them) are not co-spatial---they are located in different magnetic loops that are ``ignited'' successively by a propagating disturbance (e.g., a magnetohydrodynamic wave); consequently, the flaring loop footpoints (where the white-light emission is produced) move along the flaring arcade. The flares \#2-4 on KIC 8093473 were produced in a large active region, with the size comparable to the stellar radius (see Sections \ref{SolarSection}--\ref{BinarySection}); therefore, the distances between the different loop footpoints within the flaring arcade could be sufficient to provide a significant variation of the viewing angle.

Figure \ref{Times}(a) shows the delays between the X-ray and optical flares, defined as $\Delta t=t_0^{\mathrm{X}}-t_0^{\mathrm{WL}}$. The obtained delays were always smaller than the associated uncertainties (caused mainly by the limited time resolution of \textit{Kepler}). Nevertheless, the delays were mostly positive, which implies that the X-ray flares, as a rule, were delayed with respect to the optical ones; the only exception was a weak flare \#2 on KIC 8454353 with a relatively slow rise phase, where the peak times were poorly determined. This result is consistent with the Neupert effect \citep{Neupert1968}---a delay of flaring thermal emissions relative to nonthermal ones, which is often (but not always) observed in solar and stellar flares \citep[see, e.g.,][and references therein]{Benz2010}.

Figure \ref{Times}(b) compares the flare durations in the optical and X-ray ranges. The durations were defined as $\tau=\tau_{\mathrm{rise}}+\tau_{\mathrm{decay}}$, and their most probable values and confidence intervals were determined from the MCMC fitting procedure. The observed X-ray flares were mostly shorter than their optical counterparts (on average, $\tau^{\mathrm{WL}}/\tau^{\mathrm{X}}\sim 2$), which is inconsistent with the Neupert effect. This behaviour is uncommon for solar and stellar flares, but not extraordinary: stellar flares with $\tau^{\mathrm{WL}}/\tau^{\mathrm{X}}\gtrsim 1$ have been reported earlier, e.g., by \citet{Guarcello2019a, Guarcello2019b}. The relatively short durations of the X-ray flares can be attributed, e.g., to rapid radiative cooling of the emitting plasma, which, in turn, could be caused by a relatively high (in comparison with solar flares) plasma density in the coronal X-ray sources in the analyzed events \citep[cf.][]{Gudel2004, Mullan2006}.

\begin{figure*}
\centerline{\includegraphics{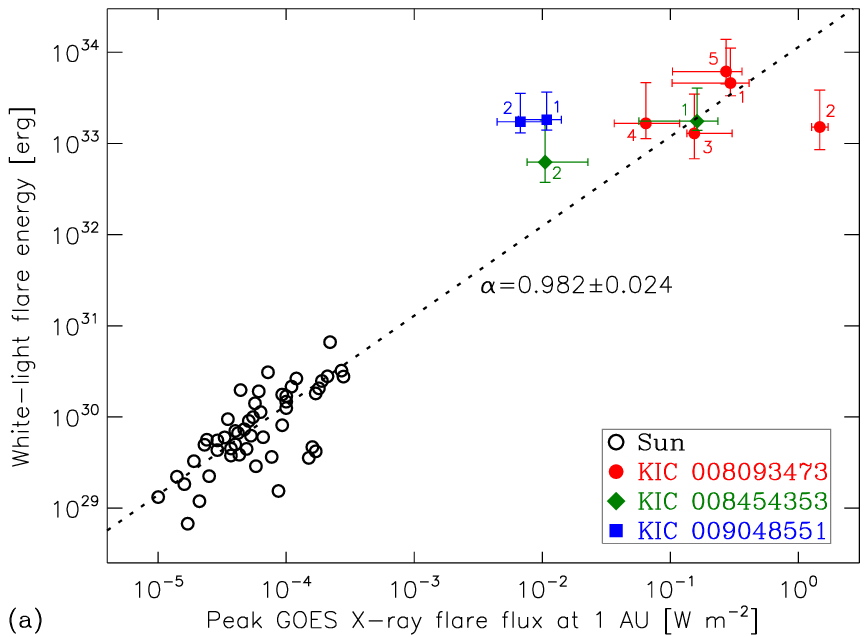}~
\includegraphics{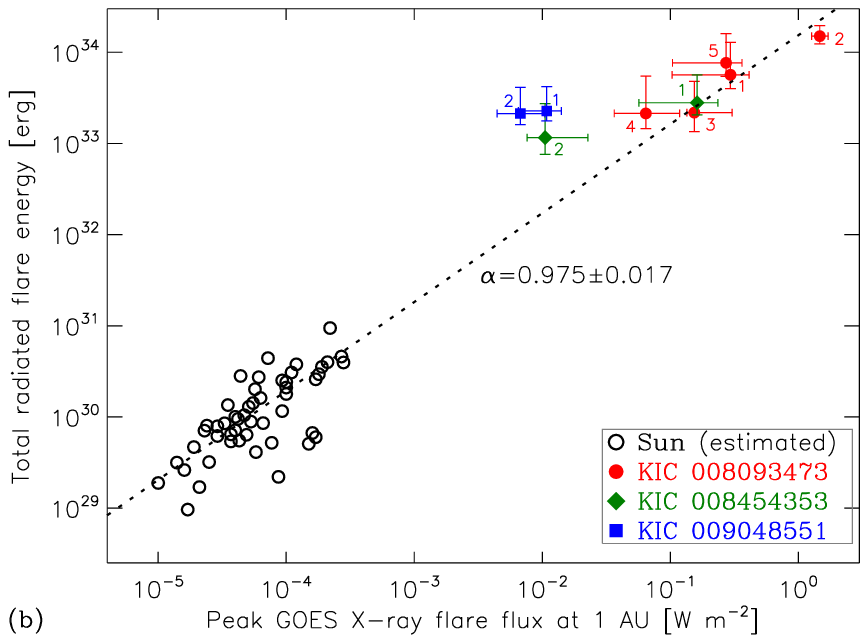}}
\caption{Flare energies vs. the peak \textit{GOES} X-ray flare fluxes for the solar \protect\citep[using the dataset from][]{Namekata2017} and stellar (this work) flares; the stellar X-ray fluxes in the \textit{GOES} energy range were estimated using the spectral fits and scaled to 1 AU distance. (a) Radiated flare energies in the white-light range. (b) Total radiated flare energies: estimated as $E_{\mathrm{flare}}=E^{\mathrm{WL}}_{\mathrm{flare}}/0.7$ for the solar flares and $E_{\mathrm{flare}}=E^{\mathrm{WL}}_{\mathrm{flare}}+E^{\mathrm{X}}_{\mathrm{flare}}$ for the stellar flares. Dotted lines represent the power-law fits. The flare numbers correspond to those in Table \protect\ref{flares}. The error bars (shown for the stellar flares only) correspond to $1\sigma$ level.}
\label{GOEScorr}
\end{figure*}

\begin{figure*}
\centerline{\includegraphics{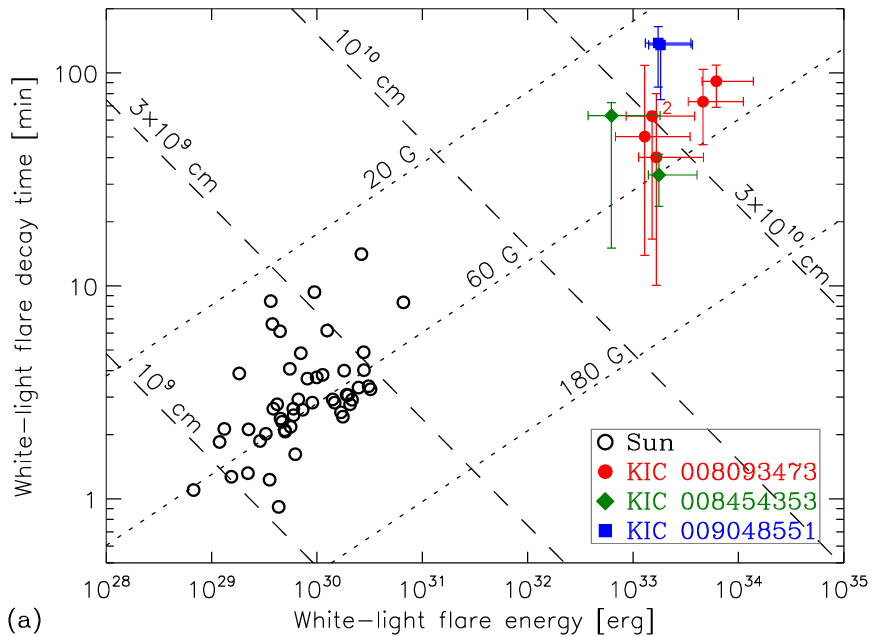}~
\includegraphics{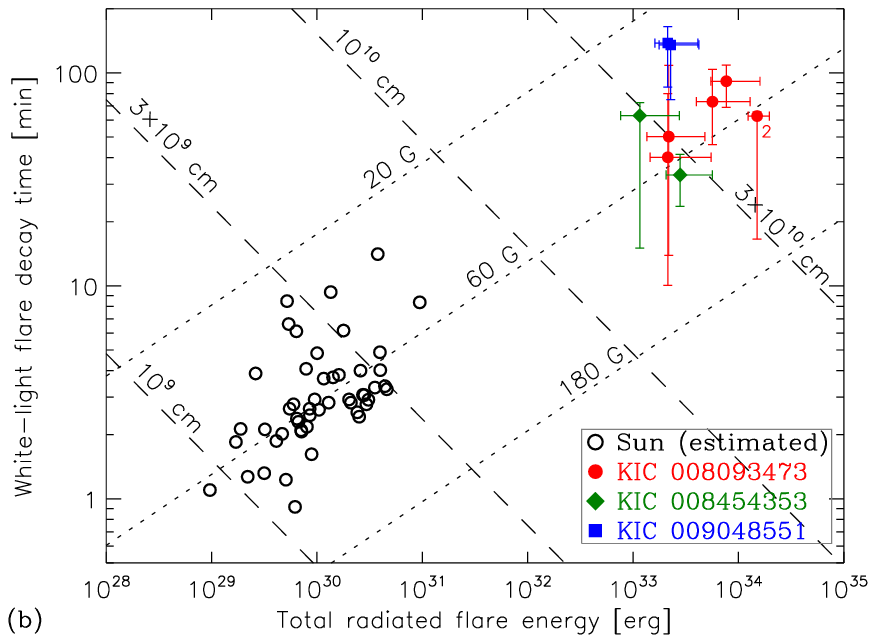}}
\caption{Flare decay times in the white-light range vs. the flare energies for the solar \protect\citep[using the dataset from][]{Namekata2017} and stellar (this work) flares. (a) Radiated flare energies in the white-light range. (b) Total radiated flare energies. The dashed and dotted lines represent the theoretical scaling laws by \protect\citet{Namekata2017}, see Equations (\protect\ref{ScalingLaws}). The flare number is shown only for the flare \#2 on KIC 8093473. The error bars (shown for the stellar flares only) correspond to $1\sigma$ level.}
\label{Scaling}
\end{figure*}

\subsection{Comparison with solar flares}\label{SolarSection}
We now compare the obtained results with the characteristics of solar flares that were also observed in both the optical and X-ray ranges; we used the sample of solar flares (50 events) presented in \citet{Namekata2017}. Figure \ref{GOEScorr}(a) shows the scatter plot of the radiated white-light flare energies vs. the peak soft X-ray fluxes in the \textit{GOES} range; we used the observed \textit{GOES} fluxes for the solar flares and the estimated equivalent \textit{GOES} fluxes for the stellar flares. We note that the distribution of stellar flares has a low-energy cutoff due to limited sensitivity of the used instruments. We fitted the relation between these parameters with a power-law dependence in the form of $E^{\mathrm{WL}}_{\mathrm{flare}}\propto (I_{\max}^{\mathit{GOES}})^{\alpha}$, which provided $\alpha=0.982\pm 0.024$. This result is consistent with conclusions and theoretical scaling laws by \citet{Namekata2017} and indicates that the flare energies, on average, are nearly proportional to the peak soft X-ray fluxes. 

In Figure \ref{GOEScorr}(b), we present estimations of the total radiated flare energies. For the stellar flares, the total flare energy was estimated as the sum of the white-light and X-ray energies: $E_{\mathrm{flare}}\simeq E^{\mathrm{WL}}_{\mathrm{flare}}+E^{\mathrm{X}}_{\mathrm{flare}}$. Since \citet{Namekata2017} did not present the radiated X-ray energies for the flares in their sample, we used the above-mentioned statistical conclusion by \citet{Kretzschmar2011} that the white-light emission is responsible for about 70\% of the total radiated energy of solar flares; therefore, the total energies of the solar flares were estimated as $E_{\mathrm{flare}}\simeq E^{\mathrm{WL}}_{\mathrm{flare}}/0.7$. Considering the total flare energy instead of the white-light one does not affect significantly the above conclusions: the fit in the form of $E_{\mathrm{flare}}\propto (I_{\max}^{\mathit{GOES}})^{\alpha}$ provided the power-law index of $\alpha=0.975\pm 0.017$. The most noticeable difference is that the flare \#2 on KIC 8093473 now agrees much better with the power-law fit; i.e., despite of an anomalously low $E^{\mathrm{WL}}_{\mathrm{flare}}/E^{\mathrm{X}}_{\mathrm{flare}}$ ratio and the largest (among the considered flares) total energy, this flare was otherwise a quite typical one. Apart from the agreement in general, six out of nine detected flares in Figure \ref{GOEScorr}(b) coincide with the obtained power-law fit within the estimated uncertainties, while three flares deviate slightly from the fit. However, in this work we do not elaborate this discrepancy, because the three outlying events were characterized by relatively low plasma temperatures ($<1$ keV, see Table \ref{xrayfits} in Appendix \ref{AppData}), and hence the extrapolated X-ray fluxes in the \textit{GOES} range could be underestimated.

\citet{Maehara2015} and \citet{Namekata2017} derived the scaling laws describing the relations between the flare parameters, under the assumptions that (a) the flare energy is proportional to the magnetic energy in the flaring volume, and (b) the flare duration is proportional to the Alfv\'en travel time through the flaring region. For a constant coronal plasma density, these scaling laws have the form:
\begin{equation}\label{ScalingLaws}
\begin{array}{c}
\tau\propto E^{1/3}B^{-5/3},\\
\tau\propto E^{-1/2}L^{5/2},
\end{array}
\end{equation}
where $\tau$ is the flare duration, $E$ is the released flare energy, $B$ is the characteristic magnetic field strength in the active region, and $L$ is the length scale of the active region\footnote{\citet{Namekata2017} estimated the length scales of solar active regions as square roots of the areas of bright flaring regions observed in EUV, and the typical coronal magnetic field strengths as $1/3$ of the average absolute values of photospheric magnetic fields within the flaring regions.}. For consistency with the results of \citet{Namekata2017}, we consider here the white-light flare decay time as an estimation of the flare duration, i.e., $\tau\simeq\tau^{\mathrm{WL}}_{\mathrm{decay}}$.

Figure \ref{Scaling} demonstrates the scatter plots of the white-light flare decay times vs. the radiated white-light or total flare energies, for the analyzed here stellar flares and the solar flares reported by \citet{Namekata2017}; theoretical lines corresponding to several constant values of $B$ and $L$, according to Equations (\ref{ScalingLaws}), are overplotted. If we consider the total radiated flare energies, the characteristic magnetic field strengths in the stellar active regions can be estimated as $B\sim 25-70$ G, which is very similar to those in the solar active regions. On the other hand, the estimated length scales of the stellar active regions ($L\sim 250\,000-500\,000$ km) far exceed those of the solar active regions. Thus, according to the $E-\tau$ diagram, the analyzed stellar superflares look like ``oversized'' versions of solar flares, with nearly the same magnetic field strengths in the reconnection sites, but much larger sizes of the corresponding active regions---comparable to the stellar radii.

\subsection{On the possible star-star interaction}\label{BinarySection}
Since two stars in our sample seem to be non-single (KIC 8454353 is likely a binary, and KIC 8093473 is either a binary or even a higher-order multiple system), we have analyzed how this multiplicity can affect the flaring processes. We have estimated the masses and radii of individual components of these systems using the empirical relations derived by \citet{Mann2015} that link the stellar mass and radius to absolute magnitude in the $K_S$ band and metallicity; we have assumed that both KIC 8093473 and KIC 8454353 are binaries consisting of two identical components each, i.e., the magnitude of an individual component $K'_S$ is related to the observed magnitude of an unresolved system $K_S$ as $K'_S=K_S+2.5\log 2$. The orbital parameters were estimated under the assumption of tidally locked binaries with circular orbits and the orbital periods equal to the rotation ones; the resulting estimations can be considered as lower limits for the orbital separations. The obtained results are presented in Table \ref{stars}; according to them, both considered systems are sufficiently separated---with the distances between the components $a$ of about $26.6$ $R_{\mathrm{star}}$ for KIC 8093473 and $12.7$ $R_{\mathrm{star}}$ for KIC 8454353. 

As follows from the previous Section, the estimated sizes of stellar active regions $L$ (i.e., heights of the flaring loops above the photospheres) are comparable to the stellar radii ($L\lesssim R_{\mathrm{star}}$) and hence much smaller than the orbital separations ($L\ll a$) for both considered systems. This implies that (a) the flaring regions are confined entirely within the coronae of the respective stars, i.e., the flaring processes occur in closed magnetic loops that belong to one of the stars, rather than in long interbinary magnetic loops potentially connecting the system components; (b) consequently, the magnetic energy released during the flares comes from the dynamo processes in the stellar interiors rather than from a star-star interaction. Therefore, although the presence of a companion can potentially provide some triggering effect (i.e., control where and when the flares occur), the flaring processes themselves are expected to be not much different from those on single stars---at least, for the flares analyzed in this work; this conclusion is supported by the correlations presented in the previous Section.

\section{Conclusion}
We have matched the databases of \textit{Kepler} and \textit{XMM-Newton}, and identified nine stellar flares (on a late K dwarf and two M dwarf systems) that were observed simultaneously in the optical and soft X-ray ranges. We have analyzed the light curves of these flares, and estimated their luminosities and total radiated energies in both spectral ranges. The main results can be summarized as follows:
\begin{itemize}
\item
In most of the analyzed flares (except one), the white-light emission dominated and was responsible for about $55-80\%$ of the total radiated energy---the energy partition similar to that in solar flares. In one event, the X-ray emission strongly dominated. The observed variations of the thermal-to-nonthermal emission ratio from flare to flare could be caused either by variations of the spectral index of nonthermal particles, or by projection effects.
\item
The X-ray flares were typically delayed after and shorter than their optical counterparts. This behaviour is consistent with the standard scenario of solar/stellar flares, but requires a faster (in comparison with the solar case) cooling of the soft X-ray-emitting plasma in the stellar flares.
\item
The solar and stellar flares seem to have a similar physical mechanism: the total flare energies are nearly proportional to the peak soft X-ray fluxes, the magnetic field strengths in the reconnection sites are confined within a relatively narrow range of values (a few tens of G), and the total flare energies are determined mainly by the sizes of the active regions. On the other hand, the estimated sizes of the stellar active regions (hundreds of thousands of km) are much larger than the sizes of solar active regions observed so far, which results in respectively higher energies of stellar flares (superflares).
\end{itemize}
Evidently, our sample of stellar flares is not representative enough; in particular, we cannot analyze the dependence of the flare parameters on the stellar parameters, and cannot explore quantitatively the effect of binarity. In the context of estimating the possibility of solar extreme events, observations of flares on the stars that are more similar to the Sun (of G class) would be of special interest. For better understanding the physical processes in flares, it would be also instructive to search for more stellar flare events with predominantly thermal emissions, analyze their occurrence rate and other characteristics, and compare them to similar ``thermal'' flares on the Sun \citep[see, e.g.,][and references therein]{Fleishman2015}. We expect that future multiwavelength observations (e.g., involving \textit{TESS}) will shed more light on the nature of stellar superflares.

\acknowledgments
This work was supported by the RFBR grant 17-52-80064 and by the Ministry of Science and Higher Education of the Russian Federation. D.Y.K. acknowledges support from the STFC consolidated grant ST/T000252/1. This research has made use of the SIMBAD database and the VizieR catalogue access tool, operated at CDS, Strasbourg, France. The authors are grateful to the referee for their constructive comments and suggestions which helped to improve the paper substantially.

\bibliographystyle{aasjournal}
\bibliography{KeplerXMM}

\appendix

\begin{figure*}
\includegraphics{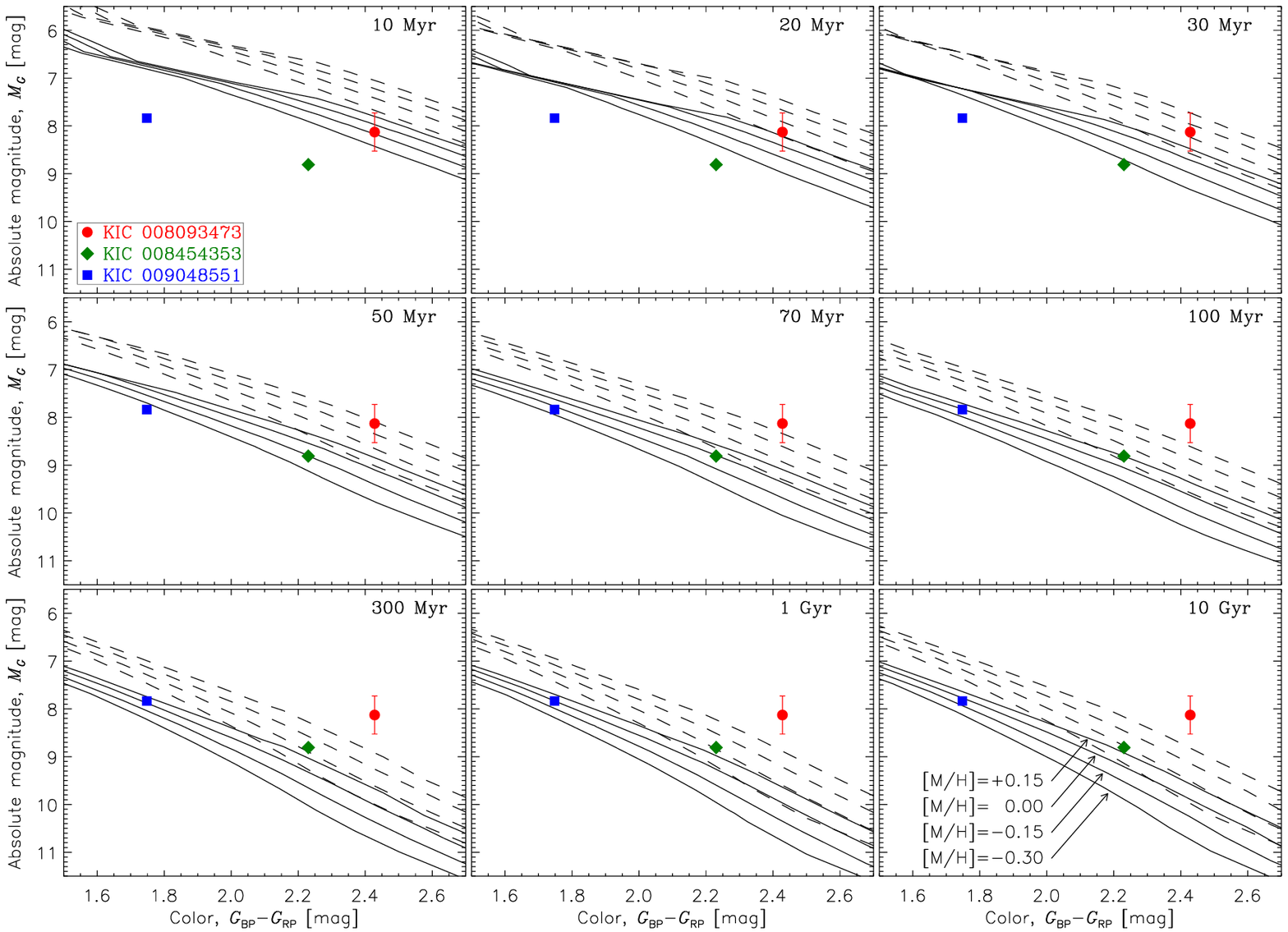}
\caption{Locations of the selected stars (see Table \protect\ref{stars}) on the Hertzsprung-Russell diagram, with the absolute stellar magnitude in the \textit{Gaia} band $M_G$ plotted vs. the \textit{Gaia} $G_{\mathrm{BP}}-G_{\mathrm{RP}}$ color. The solid lines are theoretical PARSEC isochrones \citep{Marigo2017} for main-sequence stars at different ages and metallicities [M/H]; the dashed lines represent the same isochrones shifted by $-2.5\log 2$ mag, which corresponds to unresolved binaries consisting of two identical stars. The error bars (shown for $M_G$ and KIC 8093473 only) correspond to $1\sigma$ level; other error bars are comparable to or smaller than the symbol sizes.}
\label{Isochrones}
\end{figure*}

\begin{figure*}
\includegraphics{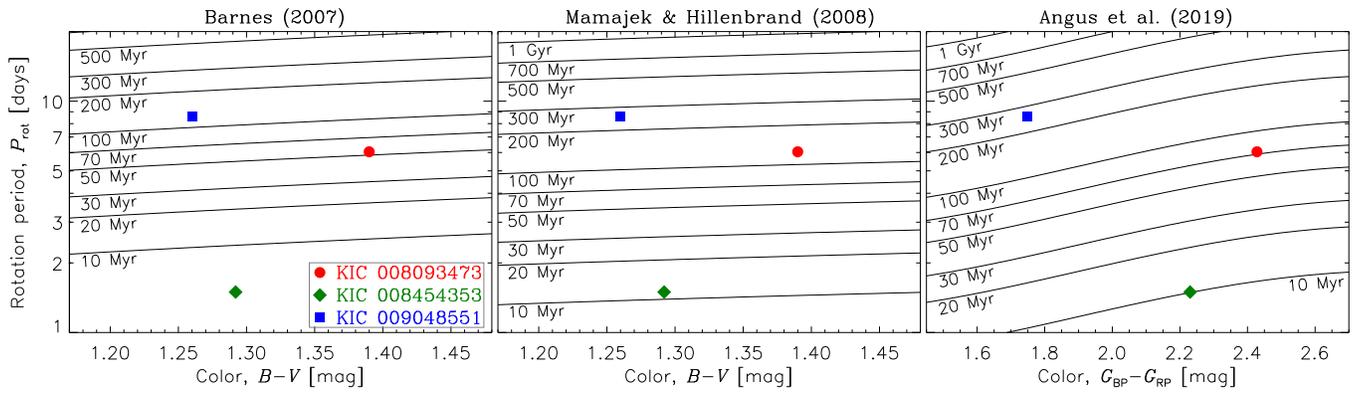}
\caption{Rotation periods of the selected stars (see Table \protect\ref{stars}) vs. their Johnson ($B-V$) or \textit{Gaia} ($G_{\mathrm{BP}}-G_{\mathrm{RP}}$) colors. The solid lines represent the empirical gyrochronology relations (isochrones) from \protect\citet{Barnes2007}, \protect\citet{Mamajek2008}, and \protect\citet{Angus2019}, at different ages. The error bars (at $1\sigma$ level) are comparable to or smaller than the symbol sizes.}
\label{Ages}
\end{figure*}

\section{Determining the stellar parameters}\label{AppStellar}
To determine the parameters and nature of the selected stars (see Table \ref{stars}), we have analyzed the available catalogs and photometry data. As said above, Figure \ref{HRdiagram} demonstrates locations of the selected stars on the Hertzsprung-Russell diagram; in addition, in Figure \ref{Isochrones} we compare the observations with the PARSEC theoretical models of stellar evolution \citep{Marigo2017}. In Figure \ref{Ages}, we compare the stellar rotation periods with empirical gyrochronology models by \protect\citet{Barnes2007}, \protect\citet{Mamajek2008}, and \protect\citet{Angus2019}; the resulting ages are consistent with those by \citet{Reinhold2015}. It follows from the analysis that:

KIC 8093473 is located well above the main sequence, but cannot be a giant or subgiant. \citet{Gaidos2016} have found for this star the photometric distance of 90 pc, which is nearly two times less than the trigonometric distance by \textit{Gaia}, i.e., the observed luminosity of KIC 8093473 is nearly four times higher than that of an average main-sequence star with the same color and metallicity. This luminosity is consistent with KIC 8093473 being either a very young single star (with an age of $\lesssim 30$ Myr, according to the PARSEC isochrones) or an unresolved binary or multiple system. However, for a single star, the age estimation from the stellar evolution models is inconsistent with the gyrochronology-derived age ($\sim 50-120$ Myr, depending on the chosen gyrochronology relation). Although the gyrochronology method becomes unreliable for the ages younger than 100 Myr \citep[][etc.]{Reinhold2015}, we conclude that KIC 8093473 is unlikely to be a single star. More likely, it is an unresolved system: e.g., a binary consisting of two similar stars with an age of $\sim 20-90$ Myr, a triple system consisting of three similar stars with an age of $\gtrsim 35$ Myr, or a quadruple system consisting of four similar stars with an age of $\gtrsim 50$ Myr; these estimations are based on comparison of the observed luminosity with the PARSEC isochrones, and are subject to uncertainties in determining the absolute stellar magnitude and metallicity. Since the gyrochronology method is not applicable to tight binaries or multiple systems, we cannot currently estimate the age of KIC 8093473 more precisely. Therefore, we conservatively conclude that this star, most likely, is an unresolved system consisting of several (from two to four) M dwarfs.

KIC 8454353 is located slightly above the main sequence. \citet{Gaidos2016} have found for this star the photometric distance of 119 pc, which is approximately $\sqrt{2}$ times less than the trigonometric distance by \textit{Gaia}, i.e., the observed luminosity of KIC 8454353 is nearly two times higher than that of an average main-sequence star with the same color and metallicity. This luminosity is consistent with KIC 8454353 being either a young single star (with an age of $\sim 35-70$ Myr, according to the PARSEC isochrones), or an unresolved binary. Like in the previous case, for a single star, the age estimation from the stellar evolution models is inconsistent with gyrochronology: the short rotation period of KIC 8454353 ($1.496$ days) is actually beyond the applicability range of the gyrochronology relations \citep{Reinhold2015}, but indicates a much younger age, while the isochrone-derived ages of $\sim 35-70$ Myr would correspond to the rotation periods of $\sim 3-6$ days. Again, although the gyrochronology method becomes unreliable for the ages younger than 100 Myr, we conclude that KIC 8454353 is unlikely to be a single star. More likely, it is an unresolved binary consisting of two similar M dwarfs with an age (based on the PARSEC isochrones) of $\gtrsim 100$ Myr. The rapid rotation of KIC 8454353 is typical of tight tidally-locked binaries \citep[e.g.,][]{Simonian2019}.

KIC 9048551 is located on the main sequence and seems to be a single K dwarf with an age (based on the PARSEC isochrones) of $\gtrsim 100$ Myr. Gyrochronology allows us to estimate the age of this star more precisely: as $\sim 120-280$ Myr, depending on the adopted gyrochronology relation.

We note that different catalogs (e.g., the \textit{Gaia} catalog and the \textit{TESS} Input Catalog) can provide considerably different parameters for some stars. In this paper, we adopt the parameters from the \textit{Gaia} catalog, because they are more suitable for our purposes. Namely, the \textit{Gaia} catalog pipeline \citep{Andrae2018} estimates the stellar effective temperature and bolometric correction from the observed color; then the absolute bolometric magnitude and the luminosity are computed using the parallax, and, finally, the stellar radius is computed using the Stefan-Boltzmann law. For an unresolved binary or multiple system, this approach provides the average effective temperature of the system components, the total luminosity of the system ($L=L_{\mathrm{A}}+L_{\mathrm{B}}+\ldots$), and the ``effective'' radius corresponding to the total visible area of the stellar disks ($R^2=R_{\mathrm{A}}^2+R_{\mathrm{B}}^2+\ldots$), i.e., the parameters needed to estimate the flare luminosity in Section \ref{WLparameters}. We have re-estimated the parameters of the stars in our sample using the above-described approach, the available photometry data, and either the relevant (i.e., color-temperature, color-bolometric correction) empirical relations from \citet{Mann2015} or the Virtual Observatory SED Analyzer\footnote{\url{http://svo2.cab.inta-csic.es/theory/vosa/}} \citep{Bayo2008}; we have obtained the parameters very similar to those in the \textit{Gaia} catalog.

On the other hand, the \textit{TESS} Input Catalog pipeline \citep{Stassun2019} uses a special procedure for the objects identified as ``cool dwarfs'': it estimates the stellar radius using the empirical relation between the radius and absolute magnitude in the $K_S$ band from \citet{Mann2015}; the absolute $K_S$ magnitude and the effective temperature are estimated using the photometry data and the \textit{Gaia} parallax. Then the stellar bolometric luminosity is computed using the radius, temperature, and Stefan-Boltzmann law. Since the magnitude-radius relation from \citet{Mann2015} is valid for single stars only, the mentioned approach provides incorrect results for unresolved binaries and multiple systems identified as cool dwarfs (including KIC 8093473 and KIC 8454353): for them, the \textit{TESS} Input Catalog underestimates the total luminosity of the system, although overestimates the luminosity and radius in comparison with those of a typical single star with the same temperature and metallicity.

\section{All parameters of the detected stellar flares}\label{AppData}
\begin{figure*}
\centerline{\includegraphics{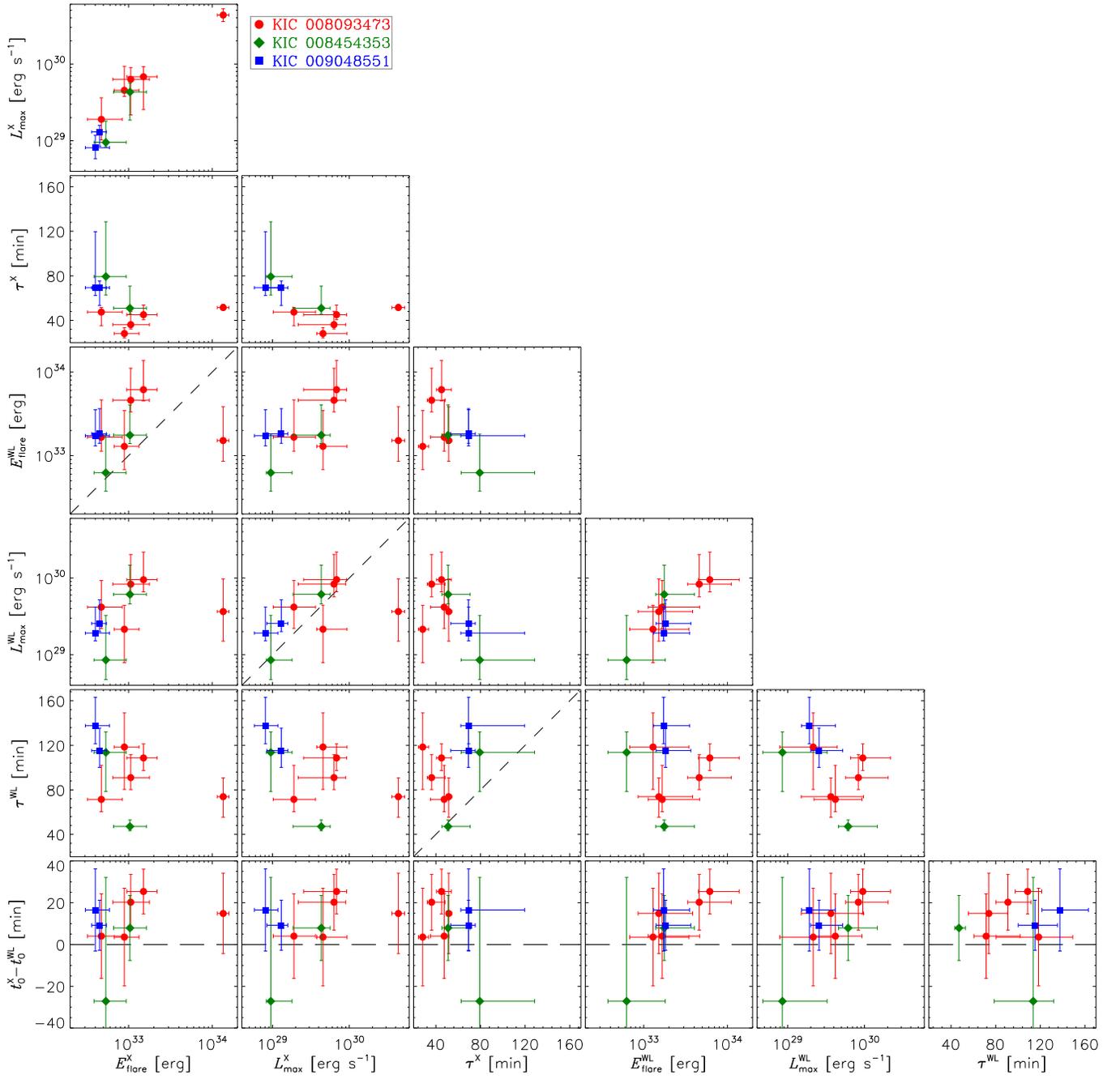}}
\caption{Scatter plots of the estimated parameters of the flares that occurred simultaneously in the X-ray (X) and white-light (WL) ranges: emitted energies ($E_{\mathrm{flare}}$), peak luminosities ($L_{\max}$), durations at $1/e$ level ($\tau$), and delays of the X-ray flares with respect to corresponding optical flares ($t_0^{\mathrm{X}}-t_0^{\mathrm{WL}}$). The error bars correspond to $1\sigma$ level.}
\label{AllCorrelations}
\end{figure*}

Table \ref{allflares} lists the parameters of all flares (i.e., all peaks in the light curves satisfying the criteria described in Section \ref{LCfitting}) detected on the considered stars. Table \ref{xrayfits} lists the X-ray spectral parameters (obtained by fitting the observed spectra with a single-temperature optically thin thermal model) for the selected ``flaring'' time intervals; each time interval can contain either one flare or several overlapping flares. Figure \ref{AllCorrelations} demonstrates the mutual correlations between various parameters of the flares that occurred simultaneously in the white-light and X-ray spectral ranges, as a corner plot.

\begin{deluxetable*}{cCCCCCCCCCCCC}
\renewcommand{\tabcolsep}{2.5pt}
\tablewidth{0pt}
\tablecaption{Parameters of all flares (with the amplitude above $1\sigma$ level and in the ``good'' time intervals) detected in the white-light (WL) and/or X-ray (X) ranges: peak times ($t_0$), rise times ($\tau_{\mathrm{rise}}$), decay times ($\tau_{\mathrm{decay}}$), peak luminosities ($L_{\max}$), emitted energies ($E_{\mathrm{flare}}$), and peak equivalent \textit{GOES} X-ray fluxes ($I_{\max}^{\mathit{GOES}}$). Only the simultaneous flares in both wavelength ranges are numbered.\label{allflares}}
\tablehead{\colhead{No.} & \colhead{$t_0^{\mathrm{WL}}$,} & \colhead{$\tau_{\mathrm{rise}}^{\mathrm{WL}}$,} & \colhead{$\tau_{\mathrm{decay}}^{\mathrm{WL}}$,} & \colhead{$L_{\max}^{\mathrm{WL}}$,} & \colhead{$E^{\mathrm{WL}}_{\mathrm{flare}}$,} & \colhead{$t_0^{\mathrm{X}}$,} & \colhead{$\tau_{\mathrm{rise}}^{\mathrm{X}}$,} & \colhead{$\tau_{\mathrm{decay}}^{\mathrm{X}}$,} & \colhead{$L_{\max}^{\mathrm{X}}$,} & \colhead{$I_{\max}^{\mathit{GOES}}$,} & \colhead{$E^{\mathrm{X}}_{\mathrm{flare}}$,}\\[-6pt]
\colhead{} & \colhead{days} & \colhead{min} & \colhead{min} & \colhead{$10^{28}$ erg~$\textrm{s}^{-1}$} & \colhead{$10^{32}$ erg} & \colhead{days} & \colhead{min} & \colhead{min} & \colhead{$10^{28}$ erg~$\textrm{s}^{-1}$} & \colhead{$10^{-2}$ W~$\textrm{m}^{-2}$} & \colhead{$10^{32}$ erg}}
\startdata
\multicolumn{12}{c}{KIC 8093473}\\
1 & 0.926_{0.917}^{0.935} &  29.6_{ 11.2}^{ 45.0} &  73.3_{ 46.1}^{104.0} &  83.2_{ 56.8}^{202.5} &  46.0_{ 33.4}^{111.1} & 0.940_{0.940}^{0.942} & 16.7_{ 9.1}^{37.6} &  19.3_{ 18.4}^{ 29.7} &  63.2_{ 21.7}^{ 89.5} &  29.5_{ 10.3}^{ 41.3} &  10.6_{  6.5}^{ 17.7}\\
\nodata & 1.035_{1.017}^{1.048} & 47.3_{12.7}^{78.6} &  66.7_{ 12.2}^{ 86.7} &  25.7_{ 10.6}^{ 54.0} &  12.2_{  6.8}^{ 33.0} & \nodata & \nodata & \nodata & \nodata & \nodata & \nodata\\
2 & 1.117_{1.107}^{1.134} &  21.9_{  7.3}^{ 57.2} &  62.7_{ 16.6}^{ 65.6} &  36.6_{ 15.0}^{ 97.0} &  15.1_{  8.6}^{ 38.4} & 1.127_{1.127}^{1.129} & 17.8_{16.9}^{21.1} &  36.8_{ 31.0}^{ 39.7} & 434.0_{357.8}^{523.7} & 147.1_{127.1}^{170.7} & 135.1_{114.8}^{157.8}\\
3 & 1.182_{1.166}^{1.199} &  32.4_{ 13.2}^{100.7} &  50.2_{ 13.9}^{108.5} &  21.5_{  7.9}^{ 44.1} &  12.9_{  6.8}^{ 34.7} & 1.185_{1.183}^{1.186} &  9.0_{ 6.9}^{23.1} &  14.6_{ 14.4}^{ 23.3} &  45.4_{ 37.8}^{ 93.3} &  15.4_{ 13.4}^{ 30.4} &   8.9_{  6.7}^{ 13.3}\\
\nodata & \nodata & \nodata & \nodata & \nodata & \nodata & 1.219_{1.218}^{1.220} & 24.7_{12.9}^{52.2} &  38.6_{ 26.8}^{ 43.3} &  42.2_{ 20.1}^{ 66.0} &  14.3_{  7.1}^{ 21.5} &  12.0_{  8.3}^{ 18.2}\\
4 & 1.276_{1.265}^{1.293} &  44.2_{ 19.8}^{ 96.7} &  40.2_{ 10.1}^{ 79.9} &  41.8_{ 22.0}^{ 92.7} &  16.6_{ 11.3}^{ 46.5} & 1.278_{1.277}^{1.279} & 29.3_{ 9.4}^{37.6} &  18.5_{ 16.4}^{ 26.6} &  19.0_{ 10.3}^{ 36.3} &   6.4_{  3.6}^{ 11.8} &   4.7_{  3.2}^{  8.4}\\
5 & 1.370_{1.364}^{1.379} &  20.4_{ 10.9}^{ 38.3} &  91.4_{ 68.9}^{108.7} &  95.0_{ 66.0}^{217.8} &  61.6_{ 45.1}^{138.1} & 1.388_{1.386}^{1.389} & 16.0_{11.8}^{31.7} &  25.5_{ 24.6}^{ 39.8} &  68.2_{ 25.4}^{ 92.5} &  27.3_{ 10.4}^{ 36.3} &  15.1_{  9.5}^{ 21.9}\\
\nodata & 1.491_{1.468}^{1.512} & 19.0_{11.7}^{98.3} &  62.8_{ 12.5}^{107.8} &  13.5_{  1.4}^{ 31.4} &   4.4_{  2.1}^{ 17.6} & \nodata & \nodata & \nodata & \nodata & \nodata & \nodata\\[3pt]
\multicolumn{12}{c}{KIC 8454353}\\
1 & 0.479_{0.477}^{0.482} &  16.9_{ 16.7}^{ 22.5} &  33.2_{ 23.6}^{ 41.4} &  61.2_{ 46.0}^{146.8} &  17.6_{ 13.9}^{ 40.4} & 0.484_{0.477}^{0.498} & 14.2_{12.4}^{41.5} &  25.9_{ 26.3}^{ 70.5} &  43.1_{ 18.5}^{ 56.3} &  16.2_{  5.7}^{ 23.5} &  10.4_{  6.6}^{ 16.3}\\
\nodata & 0.642_{0.634}^{0.647} & 13.5_{11.3}^{99.4} &  11.1_{  8.5}^{ 71.5} &   2.4_{  0.9}^{ 15.1} &   0.8_{  0.5}^{  3.6} & \nodata & \nodata & \nodata & \nodata & \nodata & \nodata\\
2 & 0.747_{0.736}^{0.750} &  77.0_{ 22.8}^{ 98.9} &  63.0_{ 15.0}^{ 72.6} &   8.6_{  4.8}^{ 32.6} &   6.2_{  3.7}^{ 18.1} & 0.728_{0.661}^{0.742} & 47.7_{13.2}^{64.7} &  26.6_{ 26.9}^{158.6} &   9.5_{  8.3}^{ 18.0} &   1.1_{  0.8}^{  2.3} &   5.3_{  3.9}^{  9.3}\\[3pt]
\multicolumn{12}{c}{KIC 9048551}\\
\nodata & 0.321_{0.318}^{0.325} & 43.0_{18.2}^{43.4} &  33.1_{  7.9}^{ 85.4} &   3.3_{  0.7}^{ 11.7} &   1.0_{  0.6}^{  3.4} & \nodata & \nodata & \nodata & \nodata & \nodata & \nodata\\
\nodata & 0.385_{0.379}^{0.386} & 40.0_{ 9.4}^{49.2} &  51.2_{ 15.0}^{114.1} &   4.9_{  1.3}^{ 13.8} &   2.1_{  1.1}^{  6.1} & \nodata & \nodata & \nodata & \nodata & \nodata & \nodata\\
\nodata & 0.465_{0.461}^{0.469} & 56.5_{23.3}^{57.9} &  91.6_{  9.8}^{111.1} &   5.3_{  1.4}^{ 12.9} &   2.2_{  1.3}^{  6.6} & \nodata & \nodata & \nodata & \nodata & \nodata & \nodata\\
1 & 0.542_{0.541}^{0.547} &  10.7_{ 10.6}^{ 20.1} & 135.2_{ 74.9}^{139.2} &  25.6_{ 20.0}^{ 52.1} &  18.2_{ 14.0}^{ 36.5} & 0.548_{0.543}^{0.559} & 12.3_{ 5.3}^{27.3} &  50.2_{ 29.1}^{ 66.6} &  13.0_{  8.5}^{ 15.9} &   1.1_{  0.6}^{  1.4} &   4.5_{  3.6}^{  5.4}\\
2 & 0.645_{0.643}^{0.655} &  13.6_{ 13.3}^{ 33.8} & 138.1_{ 85.7}^{164.9} &  19.2_{ 15.2}^{ 41.9} &  17.3_{ 13.1}^{ 35.4} & 0.657_{0.640}^{0.665} & 42.6_{14.0}^{56.2} &  36.7_{ 16.2}^{118.9} &   8.1_{  5.8}^{ 11.8} &   0.7_{  0.4}^{  1.0} &   4.0_{  3.0}^{  5.9}\\
\nodata & \nodata & \nodata & \nodata & \nodata & \nodata & 0.703_{0.696}^{0.731} & 11.0_{ 7.2}^{39.0} &  89.7_{ 27.0}^{ 93.1} &   6.3_{  4.0}^{  8.9} &   0.5_{  0.3}^{  0.8} &   3.1_{  2.1}^{  4.3}\\
\nodata & 0.770_{0.765}^{0.779} & 59.2_{17.2}^{71.8} &  54.9_{  9.7}^{143.8} &   1.3_{  0.3}^{  6.8} &   0.5_{  0.3}^{  3.1} & \nodata & \nodata & \nodata & \nodata & \nodata & \nodata\\
\nodata & 0.871_{0.865}^{0.881} & 44.6_{13.8}^{81.8} & 100.4_{ 12.9}^{162.0} &   3.9_{  0.5}^{  9.4} &   1.2_{  0.6}^{  5.2} & \nodata & \nodata & \nodata & \nodata & \nodata & \nodata\\
\nodata & 1.000_{0.987}^{1.005} & 40.7_{14.3}^{43.2} &  24.2_{  5.4}^{ 78.2} &   4.9_{  1.0}^{ 12.1} &   1.3_{  0.7}^{  3.7} & \nodata & \nodata & \nodata & \nodata & \nodata & \nodata\\
\nodata & 1.058_{1.048}^{1.067} & 20.2_{13.6}^{83.3} &  69.0_{ 28.2}^{170.1} &   3.9_{  0.5}^{  9.8} &   1.7_{  1.0}^{  6.4} & \nodata & \nodata & \nodata & \nodata & \nodata & \nodata\\
\nodata & 1.174_{1.169}^{1.190} & 42.4_{10.9}^{57.2} &  73.7_{  8.7}^{114.5} &   2.0_{  0.3}^{  6.4} &   0.4_{  0.3}^{  2.5} & \nodata & \nodata & \nodata & \nodata & \nodata & \nodata\\
\enddata
\tablecomments{The flare peak times $t_0^{\mathrm{WL}}$ and $t_0^{\mathrm{X}}$ for KIC 8093473, KIC 8454353, and KIC 9048551 are relative to MJD 55164, MJD 55829, and MJD 55735, respectively.}
\end{deluxetable*}

\begin{deluxetable*}{CcCCCC}
\tablewidth{0pt}
\tablecaption{Parameters of the X-ray spectral fits for the selected time intervals: emission measures (EM), temperatures ($T$), average luminosities ($\left<L^{\mathrm{X}}_{\mathrm{total}}\right>$) and average equivalent \textit{GOES} X-ray fluxes ($\left<I^{\mathit{GOES}}_{\mathrm{total}}\right>$). The flare numbers correspond to those in Table \protect\ref{flares}.\label{xrayfits}}
\tablehead{\colhead{Time range, days} & \colhead{Flare Nos.} & \colhead{EM, $10^{52}$ $\textrm{cm}^{-3}$} & \colhead{$T$, keV} & \colhead{$\left<L^{\mathrm{X}}_{\mathrm{total}}\right>$, $10^{28}$ erg~$\textrm{s}^{-1}$} & \colhead{$\left<I^{\mathit{GOES}}_{\mathrm{total}}\right>$, $10^{-2}$ W~$\textrm{m}^{-2}$}}
\startdata
\multicolumn{6}{c}{KIC 8093473}\\
0.911-0.961 & 1 & 20.0_{18.9}^{21.1} &  5.30_{ 4.08}^{ 6.52} & 151.3_{120.3}^{182.3} & 70.7_{57.3}^{84.0}\\
1.102-1.305 & 2, 3, 4 & 34.3_{33.6}^{35.0} &  2.31_{ 2.17}^{ 2.45} & 187.0_{164.1}^{209.8} & 63.4_{58.3}^{68.4}\\
1.355-1.442 & 5 & 19.9_{19.1}^{20.8} &  3.29_{ 2.84}^{ 3.74} & 126.3_{104.2}^{148.3} & 50.5_{42.7}^{58.3}\\[3pt]
\multicolumn{6}{c}{KIC 8454353}\\
0.472-0.562 & 1 & 4.1_{ 3.7}^{ 4.5} &  2.84_{ 1.91}^{ 3.78} &  24.3_{ 18.2}^{ 30.4} &  9.1_{ 5.6}^{12.7}\\
0.666-0.771 & 2 & 3.7_{ 3.3}^{ 4.1} &  0.76_{ 0.69}^{ 0.82} &  14.2_{ 12.6}^{ 15.8} &  1.6_{ 1.1}^{ 2.0}\\[3pt]
\multicolumn{6}{c}{KIC 9048551}\\
0.528-0.725 & 1, 2 & 2.7_{ 2.5}^{ 2.8} &  0.66_{ 0.64}^{ 0.69} &  10.3_{  9.6}^{ 10.9} &  0.9_{ 0.7}^{ 1.0}\\
\enddata
\tablecomments{The time ranges for KIC 8093473, KIC 8454353, and KIC 9048551 are relative to MJD 55164, MJD 55829, and MJD 55735, respectively.}
\end{deluxetable*}

\end{document}